\documentclass[10pt,showkeys,aps,prc,twocolumn,floatfix]{revtex4-1}

\usepackage{graphicx} 
\usepackage{hyperref}  
\usepackage{amssymb}  
\usepackage{amsmath}

\begin{document}
\title{How Analytic Choices Can Affect the Extraction of Electromagnetic Form Factors
       from Elastic Electron Scattering Cross Section Data}

\author{Scott~K.~Barcus}
\author{Douglas~W.~Higinbotham}
\author{Randall~E.~McClellan}
\affiliation{Jefferson Lab, Newport News, VA 23606}

\begin{abstract}
Scientists often try to incorporate prior knowledge into their regression algorithms, such 
as a particular analytic behavior or a known value at a kinematic endpoint. 
Unfortunately, there is often no unique way to make use of this prior knowledge, 
and thus, different analytic choices can lead to very different regression results 
from the same set of data. 
To illustrate this point in the context of the proton electromagnetic form factors,
we use the Mainz elastic data with its 1422 cross section points and 31 normalization parameters. 
Starting with a complex unbound non-linear regression, we will show how the addition of 
a single theory-motivated constraint removes an oscillation from the magnetic form factor 
and shifts the extracted proton charge radius.   
We then repeat both regressions using the same algorithm, but with a rebinned version of the Mainz dataset.
These examples illustrate how analytic choices, such as the function that is being used or 
even the binning of the data, can dramatically affect the results of a complex regression.
These results also demonstrate why it is critical when using regression algorithms 
to have either a physical model in mind or a firm mathematical 
basis.
\end{abstract}

\maketitle

\section{Introduction}

Silberzahn~{\it{et al.}}~\cite{Silberzahn:2018} points out that there is often little appreciation for how different analytic strategies can affect a reported result. In this work, we illustrate how analytic choices can impact the extraction of the electromagnetic form factors and the associated charge radii from electron scattering data. These extractions are frequently done with complex non-linear regression algorithms and tend to make use of prior information about the limiting behavior of the electromagnetic form factors
to help constrain the value of experimental normalization parameters.
Also, while many tend to look at all regressions as being the same, in fact there are 
different types of regressions such as descriptive, predictive, and explanatory.

A descriptive model is used to capture the features of a dataset in a compact manner without reliance on an underlying theory.  A predictive model is any statistical model which tries to generalize beyond the data that is being fitted. Finally, explanatory modeling takes a theory based model and tests that model's hypothesis by applying it to data. Further details about these differences can be found in Ref.~\cite{shmueli2010}. Though the type of regression model being developed is not always clearly stated, it is yet another choice that affects how scientists design their regression algorithms.

\section{Proton Elastic Scattering}

There has been renewed interest in proton elastic scattering data due to muonic hydrogen Lamb shift results which determined the charge radius of the proton to be 0.84078(39)~fm~\cite{Pohl:2010zza,Antognini:1900ns}, a result in stark contrast to the CODATA-2014 recommended value of 0.8751(61)~fm~\cite{Mohr:2015ccw}. This systematic difference was known as the proton radius puzzle~\cite{Pohl:2013yb,Carlson:2015jba,Miller:2018zfu}.  We will show that determining the proton's charge radius is highly dependent on the analytic choices made when selecting a model to describe the world data. 

In the plane-wave Born approximation, the cross section for elastic electron scattering on a proton is given by:

\begin{align}
	\sigma  = & \sigma_\mathrm{Mott} \times \nonumber \\
 	& \left[\frac{G_E^2\left(Q^2\right)+\tau G_M^2\left(Q^2\right)}{1+\tau}+2\tau 							G_M^2\left(Q^2\right)\tan^2\left(\frac{\theta}{2}\right)\right] \text{,}
\label{eqrosen}
\end{align}

\noindent where $\sigma_\mathrm{Mott}$ is the Mott cross section, $G_E$ and $G_M$ are the electric and magnetic Sachs form factors respectively, $\tau = \frac{Q^2}{4 m_p^2}$, $Q^2=4E_{\mathrm{Beam}}E^\prime\sin^2\left(\frac{\theta}{2}\right)$, $E_{\mathrm{Beam}}$ is the energy of the electron beam, $E'$ is the energy of the outgoing electron, $\theta$ is the scattering angle of the outgoing electron, and $m_p$ is the mass of the proton.

The proton charge radius, $r_p$, is extracted from the cross sections by determining the slope of the electric form factor, $G_E$, in the limit of four-moment transfer, $Q^2$, approaching zero~\cite{Miller:2018ybm}: 

\begin{equation}
\label{eq:radius}
  r_p \equiv 
    \left( -6  \left. \frac{dG_E(Q^2)}{dQ^2}
    \right|_{Q^{2}=0} \right)^{1/2} \>.
\end{equation}

\noindent Since the scattering data is measured at finite $Q^2$, an extrapolation is required to extract the charge radius. A purely mathematical fit to the scattering data would be a descriptive model and would generally only be valid in the region of the data making extrapolation risky. When extrapolating back to $Q^2$=0 it is desirable to use a predictive model. This requires extra care such as adding physics considerations to the model and/or mathematical requirements to keep the fit well behaved (e.g. not diverging in the region) and not unduly complex. Authors have taken many different approaches to this extraction, yielding different
outcomes~\cite{Horbatsch:2015qda,Griffioen:2015hta,Higinbotham:2015rja,Lee:2015jqa,Graczyk:2014lba,
Lorenz:2014vha,Horbatsch:2016ilr,Alarcon:2018zbz,Alarcon:2020kcz,Zhou:2018bon,Mihovilovic:2020dmd}.

\section{Regressions}

To illustrate how analytic choice can strongly affect the extracted radius, we first use the Mainz dataset of 1422 cross section points along with its 31 normalization parameters~\cite{Bernauer:2010wm}. As noted in the work of Bernauer {\it{et al.}}~\cite{Bernauer:2013tpr}, knowledge of the absolute value of cross sections is limited by the determination of the absolute luminosity, which in turn is limited by the uncertainty of the target thickness and beam current. 
In order to compensate for these uncertainties, 
normalization parameters were introduced to the original fits of this data which were constrained only by the value of the charge and magnetic form factors in the limit of $Q^2=0$.   While there is no debate about the value of the form factors at $Q^2=0$, how the data at
finite $Q^2$ is connected to the known end-point brings a model dependence to the analysis that is not easily understood. 

These parameters are taken in combinations to link sets of data together, with the final value of each cross section point defined by:

\begin{equation}
\label{eq:sigma_exp}
\sigma_{exp} = \sigma_{p} \cdot normA_p \cdot normB_p,
\end{equation}

\noindent where $normA_p$ and $normB_p$ are the two normalization parameters associated with that data point. A complete list of the 31 different normalization parameters, N$_j$, that are taken in 34 unique combinations for the 1422 points, is shown in Table~\ref{table:norm}. Further details of how these parameters connect to each of the 1422 cross section points can be found in the supplemental material of Bernauer {\it{et al.}}~\cite{Bernauer:2013tpr}.

For our example of how analytic choice can effect the outcome, we first fit the Mainz dataset with an unbound complex non-linear regression 
with the form factors parameterized in terms of polynomials:

\begin{align}
\label{eq:polynomial}
G_{E,\mathrm{polynomial}}(a^{E}_i,Q^2) & = 1+\sum_{i=1}^n a^{E}_i Q^{2\, i} \;\; \text{and}\\
\label{eq:polynomial2}
G_{M,\mathrm{polynomial}}(a^{M}_i,Q^2) & = \mu_p \left( 1+\sum_{i=1}^n a^{M}_i Q^{2\, i} \right) \text{,}
\end{align}

\noindent where $\mu_p$ is the magnetic moment of the proton and $n$ is the order of the polynomial.
This is one of the many form factors parameterizations described in Ref.~\cite{Bernauer:2013tpr} that has been used for extracting the proton radius from the Mainz dataset.  

For these regressions, we perform a weighted least squares minimization with a $\chi^2$ function defined as follows:

\begin{equation}
\chi^2 = \sum_{p=1}^{p_{max}} \left( \frac{ \sigma_{Model}(E_p,\theta_p) - \sigma_p \cdot normA_p \cdot normB_p }{ \Delta\sigma_{p} \cdot normA_p \cdot normB_p} \right)^2 \text{,}
\end{equation}

\noindent where for each data point $p$ there is a cross-section, $\sigma_p$, with energy $E_p$, angle $\theta_p$, 
and normalization parameters $normA$ and $normB$ as shown in Table~\ref{table:norm}.   
As was done in the Mainz fits, the normalization parameters are allowed to float freely.

We repeat this same regression adding one requirement: 
that the terms of the polynomial have successively alternating signs.   This makes the polynomial 
more closely resemble a ``completely monotone'' function and is referred to as a bound regression.  
A true completely monotone function, $f$, would possess derivatives, $f^n$, of all orders such that $(-1)^n f^n(x) \geq 0 , x > 0$~\cite{Merkle:2012}.   
Of course for a finite order polynomial being fit to experimental data, we are simply approximating 
a completely monotone function over the range of the data by alternating the signs of the terms.
This seemingly simple constraint imposes an analytic behavior to the form factors that is 
constant with nuclear physics calculations
such as chiral effective field theory~\cite{Alarcon:2018irp}.
In statistics terms, 
adding the condition that the polynomial approximate a physically motivated function would be classified as creating a 
robust regression model~\cite{Recipes:2007}. Robust regression models are designed
such that they are not
unduly affected by outliers, whereas least squares estimates are highly sensitive to outlying points 
as illustrated in Appendix~\ref{sec:robust}. 

As a further check of how sensitive these two functions are to the handling of the data in the fit, we use the rebinned version of the Mainz data that is provided in the supplemental material of Ref.~\cite{Lee:2015jqa}. These authors carefully rebinned and re-weighted the full Mainz dataset and provided a new set of 658 cross section points, though with the same 31 normalization parameters as the original set. By simply replacing the original Mainz dataset with this set, we can repeat our unbound and bound regressions. 

While regressions that are linear in terms can be solved exactly, this is not the case with non-linear 
regressions where algorithms can converge in a local or non-physical minimum; thus choosing reasonable 
initialization parameters is an important step when developing non-linear regression algorithms. To have reasonable initialization parameters for our complex non-linear regressions, we first perform a regression with dipole functions for $G_E$ and $G_M$ and use the resultant normalization parameters as initialization parameters in the more complex regressions.

\begin{table}[htb]
\caption{The 34 different combinations of the 31 normalization parameters, N$_j$, found in Ref.~\cite{Bernauer:2013tpr} which link the data together along with the number of data points and the $Q^2$ range of each dataset.}
\begin{tabular}{rccccl}
Energy  & Spec. & $normA$ & $normB$ & Points& $Q^2$ Range [GeV$^2$] \\ \hline 
180 MeV & B     & N$_1$ & N$_3$ & 106	& 0.0038 -- 0.0129 \\
        & B     & N$_1$ & N$_4$ & 41	& 0.0101 -- 0.0190 \\
        & A     & N$_3$ & -     & 102	& 0.0112 -- 0.0658 \\
        & B     & N$_1$ & N$_5$ & 19	& 0.0190 -- 0.0295 \\
        & C     & N$_2$ & N$_4$ & 38    & 0.0421 -- 0.0740 \\
        & C     & N$_2$ & N$_5$ & 17    & 0.0740 -- 0.0834 \\
\hline
315 MeV & B  	& N$_6$	& N$_9$	& 104	& 0.0111 -- 0.0489 \\
	& A  	& N$_7$	& N$_9$	& 38    & 0.0430 -- 0.1391 \\
	& A  	& N$_9$	& -	& 40    & 0.0479 -- 0.1441 \\ 
	& C  	& N$_8$	& N$_9$	& 62	& 0.1128 -- 0.2131 \\ 
\hline
450 MeV & B  	& N$_{10}$& N$_{13}$& 77	& 0.0152 -- 0.0572 \\
	& B  	& N$_{10}$& N$_{15}$& 52        & 0.0572 -- 0.1175 \\
	& A  	& N$_{13}$& - 	    & 42	& 0.0586 -- 0.2663 \\
	& B  	& N$_{10}$& N$_{14}$& 17	& 0.0589 -- 0.0851 \\
	& A  	& N$_{11}$& N$_{13}$& 36	& 0.0670 -- 0.2744 \\ 
	& C  	& N$_{12}$& N$_{15}$& 50	& 0.2127 -- 0.3767 \\
	& A  	& N$_{14}$& -	    & 2         & 0.2744 -- 0.2744 \\
\hline
585 MeV & B	& N$_{16}$& N$_{18}$& 41 	& 0.0255 -- 0.0433 \\
	& B	& N$_{16}$& N$_{19}$& 47	& 0.0433 -- 0.1110 \\
	& A	& N$_{18}$& -	    & 27	& 0.0590 -- 0.0964 \\
	& B	& N$_{16}$& N$_{20}$& 21 	& 0.0920 -- 0.1845 \\
	& A	& N$_{19}$& - 	    & 37	& 0.0964 -- 0.4222 \\ 
	& C	& N$_{17}$& N$_{20}$& 20	& 0.3340 -- 0.5665 \\
\hline
720 MeV & B	& N$_{21}$& N$_{25}$& 47	& 0.0711 -- 0.1564 \\
	& A	& N$_{25}$& -	    & 46	& 0.1835 -- 0.6761 \\
	& C	& N$_{24}$& N$_{26}$& 28	& 0.6536 -- 0.7603 \\
	& B	& N$_{23}$& N$_{26}$& 27	& 0.2011 -- 0.2520 \\
	& C	& N$_{22}$& N$_{26}$& 37	& 0.4729 -- 0.7474 \\
	& B	& N$_{21}$& N$_{26}$& 36 	& 0.1294 -- 0.2435 \\ 
\hline
855 MeV	& B	& N$_{27}$& N$_{31}$& 35	& 0.3263 -- 0.4378 \\
	& C	& N$_{28}$& N$_{31}$& 31 	& 0.7300 -- 0.9772 \\
	& A	& N$_{29}$& N$_{30}$& 32 	& 0.3069 -- 0.5011 \\
	& A	& N$_{29}$& -	    & 13	& 0.5274 -- 0.7656 \\
	& B	& N$_{27}$& N$_{29}$& 54 	& 0.0868 -- 0.3263 \\
\hline
\end{tabular}
\label{table:norm}
\end{table}

\begin{figure*}[p]
\centering
\begin{minipage}[b]{0.48\textwidth}
\includegraphics[width=0.99\columnwidth]{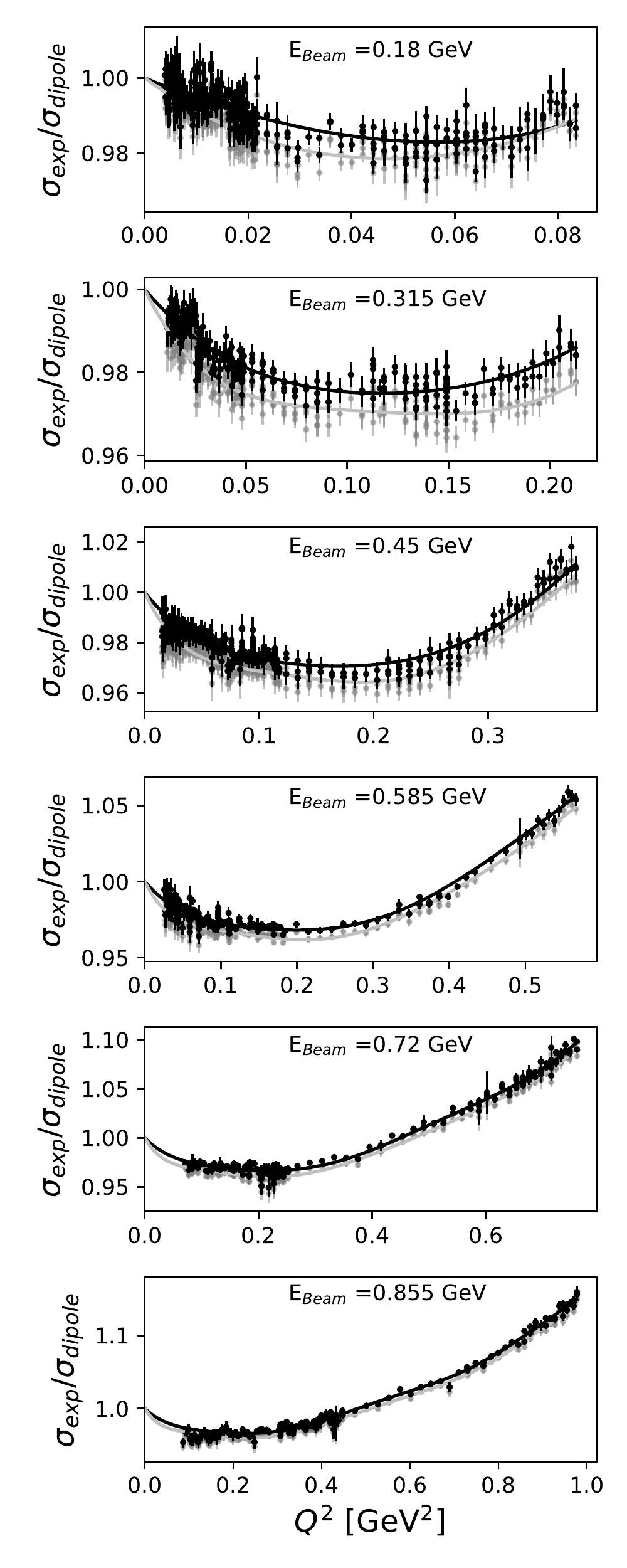}
\end{minipage}\qquad
\begin{minipage}[b]{0.48\textwidth}
\includegraphics[width=0.99\columnwidth]{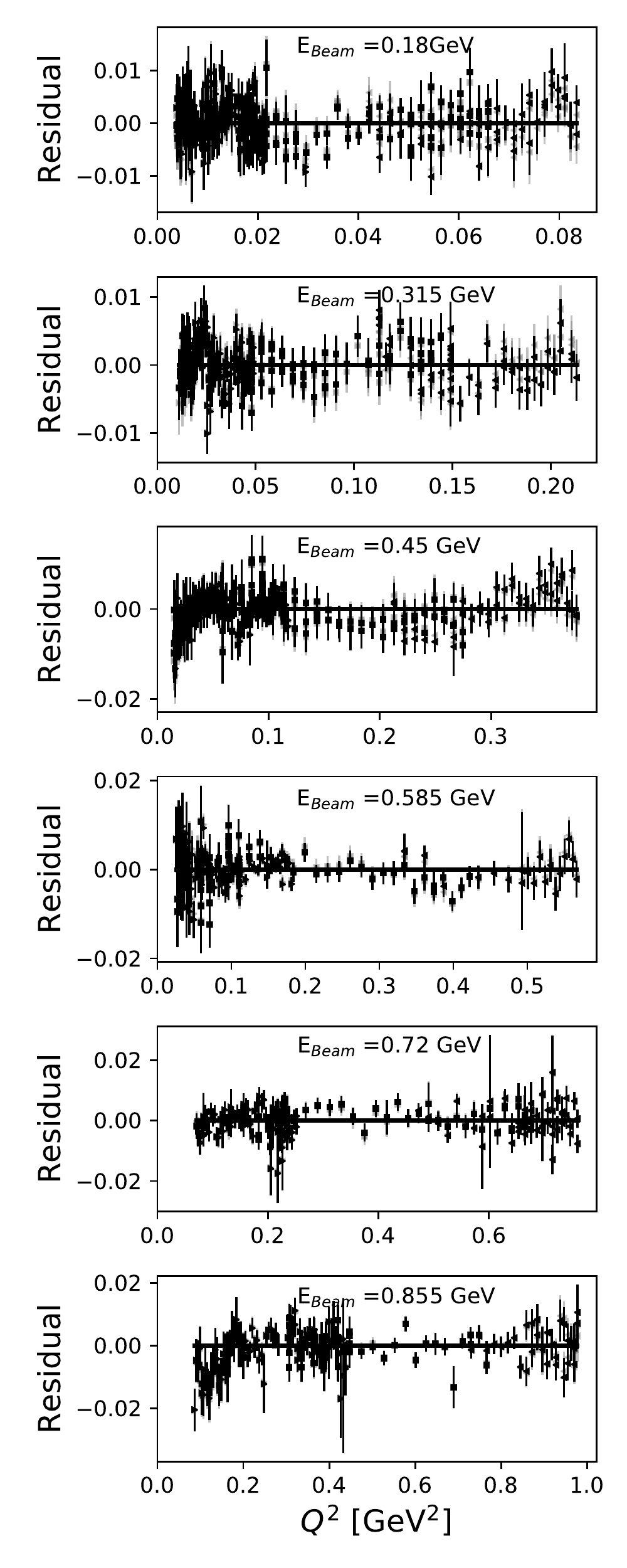}
\end{minipage}
    \caption{The 1422 Mainz cross section points plotted vs. $Q^2$ for the six different incident beam energies and fit with unbound and bound polynomials. The gray points were analyzed using an unbound eleventh order polynomial regression in $G_E$ and $G_M$ while the black points used a bound eleventh order polynomial regression constrained to alternate term signs. The systematic difference in the location of the points is due to how the 31 normalization parameters in the fit change based on the choice of using either the unbound or bound functions in the regression. While the mean values are clearly different for these fits, the residuals of the fits to their respective functions are quite similar.}
\label{fig:fit1422}
\end{figure*}

\begin{figure*}[p]
\centering
\begin{minipage}[b]{0.48\textwidth}
\includegraphics[width=0.99\columnwidth]{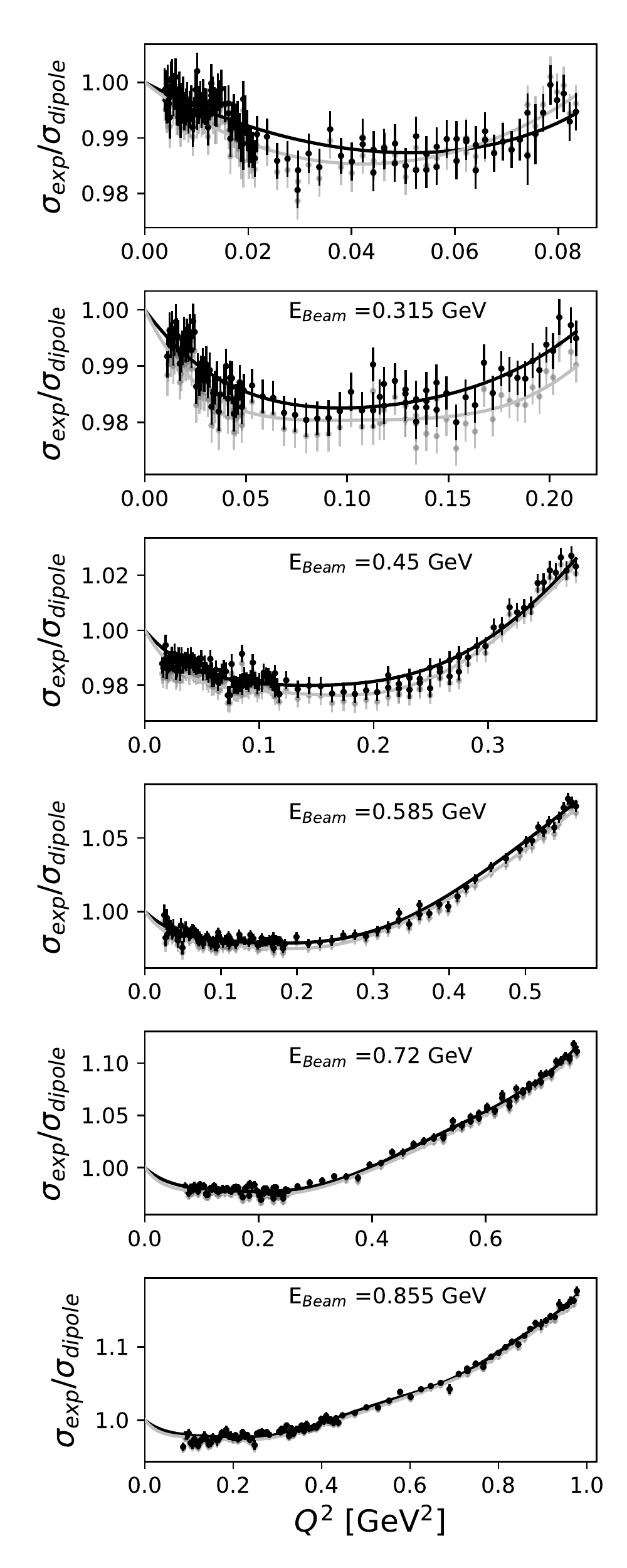}
\end{minipage}\qquad
\begin{minipage}[b]{0.48\textwidth}
\includegraphics[width=0.99\columnwidth]{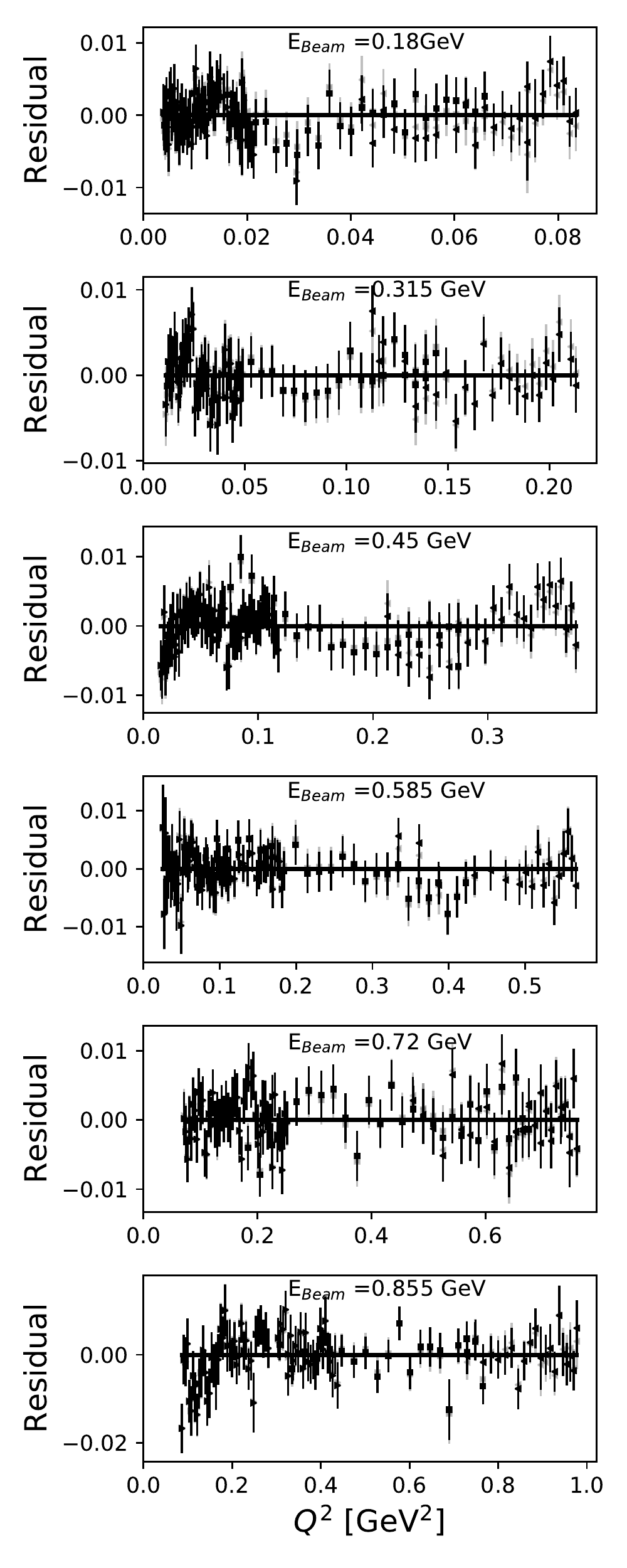}
\end{minipage}
    \caption{The 658 rebinned Mainz cross section points plotted vs. $Q^2$ for the six different incident beam energies and fit with unbound and bound polynomials. The gray points were analyzed using an unbound eleventh order polynomial regression in $G_E$ and $G_M$ while the black points used a bound eleventh order polynomial regression constrained to alternate term signs. The systematic difference in the location of the points is due to how the 31 normalization parameters in the fit change based on the choice of using either the unbound or bound functions in the regression. While the mean values again differ for the two fits, the residuals of the fits to their respective functions are quite similar.}
\label{fig:fit658}
\end{figure*}

In Table~\ref{poly:madness} and Table~\ref{table:ff658} we show the results of fitting with both the unbound and bound regressions for the 1422 Mainz cross section points and the rebinned 658 Mainz cross section points respectively. The regression results and residuals are shown graphically in Fig.~\ref{fig:fit1422} and Fig.~\ref{fig:fit658} for the 1422 Mainz cross section points and the rebinned 658 Mainz cross section points respectively. For clarity we divide $\sigma_{exp}$ by $\sigma_{dipole}$, where $\sigma_{dipole}$ is simply Eq.~\ref{eqrosen} with standard dipole form factors:
\begin{align}
G_{E,\mathrm{dipole}}(Q^2) & = \left( 1+\frac{Q^2}{0.71~\mathrm{GeV}^2} \right)^{-2} \; \text{and} \\ 
G_{M,\mathrm{dipole}}(Q^2) & = \mu_p \left( 1+\frac{Q^2}{0.71~\mathrm{GeV}^2} \right)^{-2}.
\end{align}
The fits of the unbound and bound regressions clearly differ significantly for both the original and the rebinned Mainz cross section points, but the residuals of each fit are quite similar. The reason the locations of the cross section data points differ between the unbound and bound regressions in Fig.~\ref{fig:fit1422} and Fig.~\ref{fig:fit658} is due to the choice of regression model shifting the 31 normalization parameters to maintain agreement with our prior knowledge of the values of the electromagnetic form factors in the limit of $Q^2$ = 0. The magnitude of individual normalizaton shifts is a few tenths of a percent, which is much smaller than absolute normalizatons can be determined, but can have a clear effect on the results as the point-to-point uncertainties are also just few tenths of a percent.  


\begin{table}[htb]
\caption{The values of the polynomial terms for the unbound and bound regressions of the 1422 cross section points following the notation of Eq.~\ref{eq:polynomial} and~\ref{eq:polynomial2}. If one wishes to interpret the charge form factor slope term $(i=1)$ in terms of charge radius using Eq.~\ref{eq:radius}, one finds the unbound fit gives a charge radius of 0.882~fm while the bound fit gives a charge radius of 0.854~fm.}
\begin{tabular}{c |cc | cc}
  & \multicolumn{2}{c}{unbound}& \multicolumn{2}{c}{bound}\\
i & $a^E_i$ & $a^M_i$ & $a^E_i$ & $a^M_i$ \\ \hline
1 & -3.331 & -2.523 & -3.124 & -2.800 \\
2 & 13.05 & -0.7081 & 8.821 & 5.188 \\
3 & -63.68 & 40.16 & -25.74 & -5.742 \\
4 & 249.4 & -176.7 & 60.06 & 2.806 \\
5 & -658.6 & 380.3 & -89.41 & -0.000 \\
6 & 1099 & -392.6 & 72.48 & 0.01034 \\
7 & -987.6 & 11.53 & -24.23 & -0.2766 \\
8 & 57.38 & 442.4 & 0.0000 & 0.0000 \\
9 & 853.4 & -492.1 & -0.0061 & -0.0009 \\
10 & -810.5 & 230.3 & 0.0081 & 0.0013 \\
11 & 250.4 & -40.92 & -0.0000 & -0.0000 \\
\end{tabular}
\label{poly:madness}
\end{table}

\begin{table}[htb]
\caption{The values of the polynomial terms for the unbound and bound regressions of the 658 cross section points of rebinned data~\cite{Lee:2015jqa} following the notation of Eq.~\ref{eq:polynomial} and~\ref{eq:polynomial2}. If one wishes to interpret the charge form factor slope term $(i=1)$ in terms of charge radius using Eq.~\ref{eq:radius}, one finds the unbound fit gives a charge radius of  0.863~fm while the bound fit gives a radius of 0.845~fm.}
\begin{tabular}{c |cc | cc}
  & \multicolumn{2}{c}{unbound}& \multicolumn{2}{c}{bound}\\
i & $a^E_i$ & $a^M_i$ & $a^E_i$ & $a^M_i$ \\ \hline
1 & -3.191 & -2.465 & -3.061 & -2.760 \\
2 & 10.83 & -0.7271 & 8.413 & 4.979 \\
3 & -44.59 & 35.32 & -24.46 & -5.196 \\
4 & 157.2 & -136.4 & 58.23 & 2.193 \\
5 & -404.2 & 228.8 & -89.36 & -0.000 \\
6 & 712.6 & -98.11 & 74.77 & 0.5035 \\
7 & -733.1 & -234.1 & -25.73 & -0.5330 \\
8 & 133.4 & 349.9 & 0.0000  & 0.0000 \\
9 & 632.8 & -122.4 & -0.0000 & -0.0000 \\
10 & -695.5 & -56.49 & 0.0000 & 0.0000 \\
11 & 232.9 & 35.80 & -0.0000 & -0.0000 \\
\end{tabular}
\label{table:ff658}
\end{table}

\section{Model Selection}

\begin{figure}[thb]
\includegraphics[width=\columnwidth]{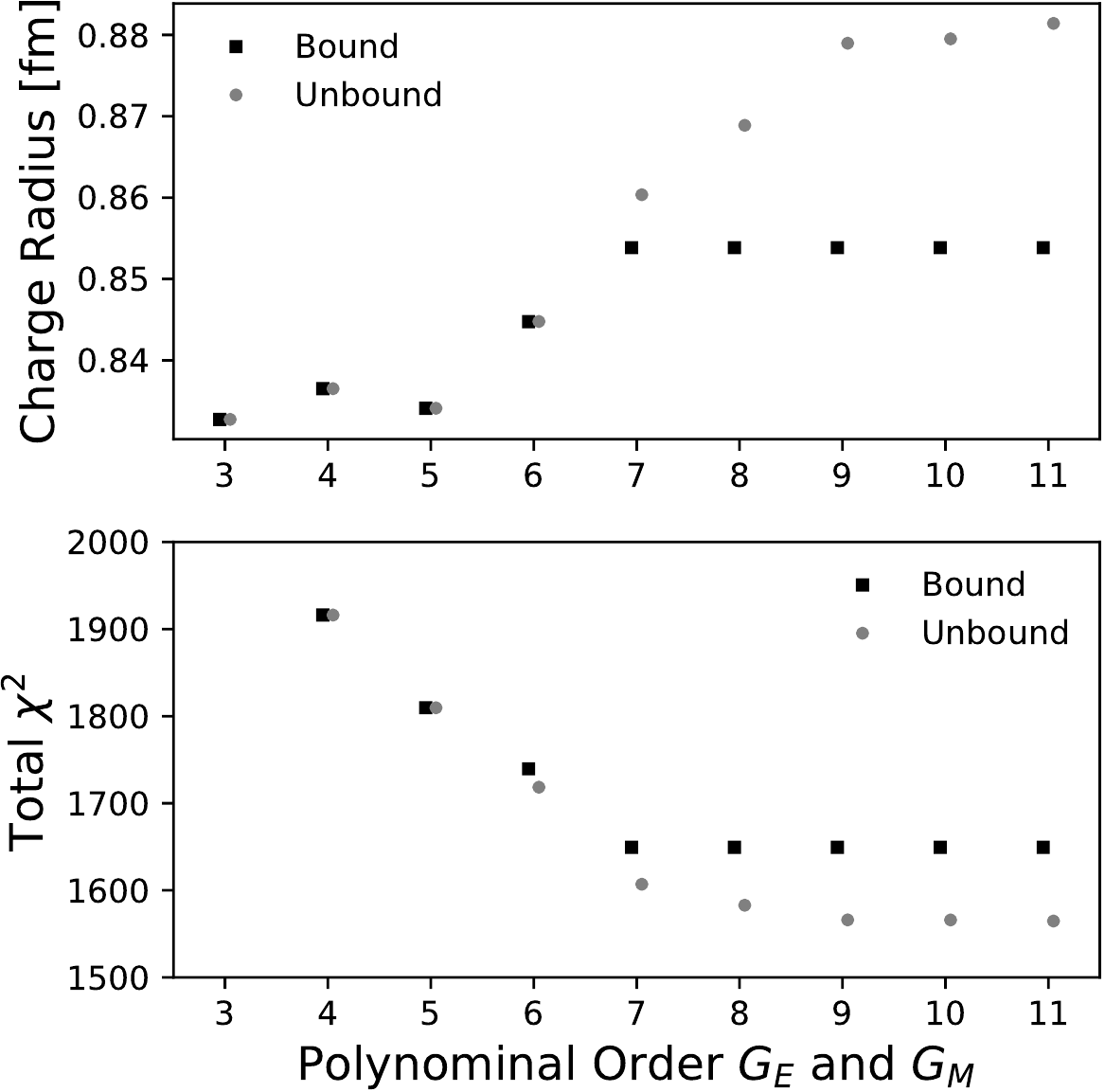}
\caption{Total $\chi^2$ vs. the number of fit parameters in $G_E$ and $G_M$ using Eq.~\ref{eq:polynomial} and~\ref{eq:polynomial2} for both the unbound and bound polynomial regressions of the full 1422 point Mainz dataset. Total $\chi^2$ will decrease as parameters are added, but at some point no significant improvement will be made where significance is defined using the statistical criteria.  With both AIC and BIC, the most appropriate bound fits are the $7^{th}$ order while for the unbound descriptive fits the most appropriate order is $10^{th}$.}
\label{TheChi2}
\end{figure}

\begin{figure}[htb]
\includegraphics[width=\columnwidth]{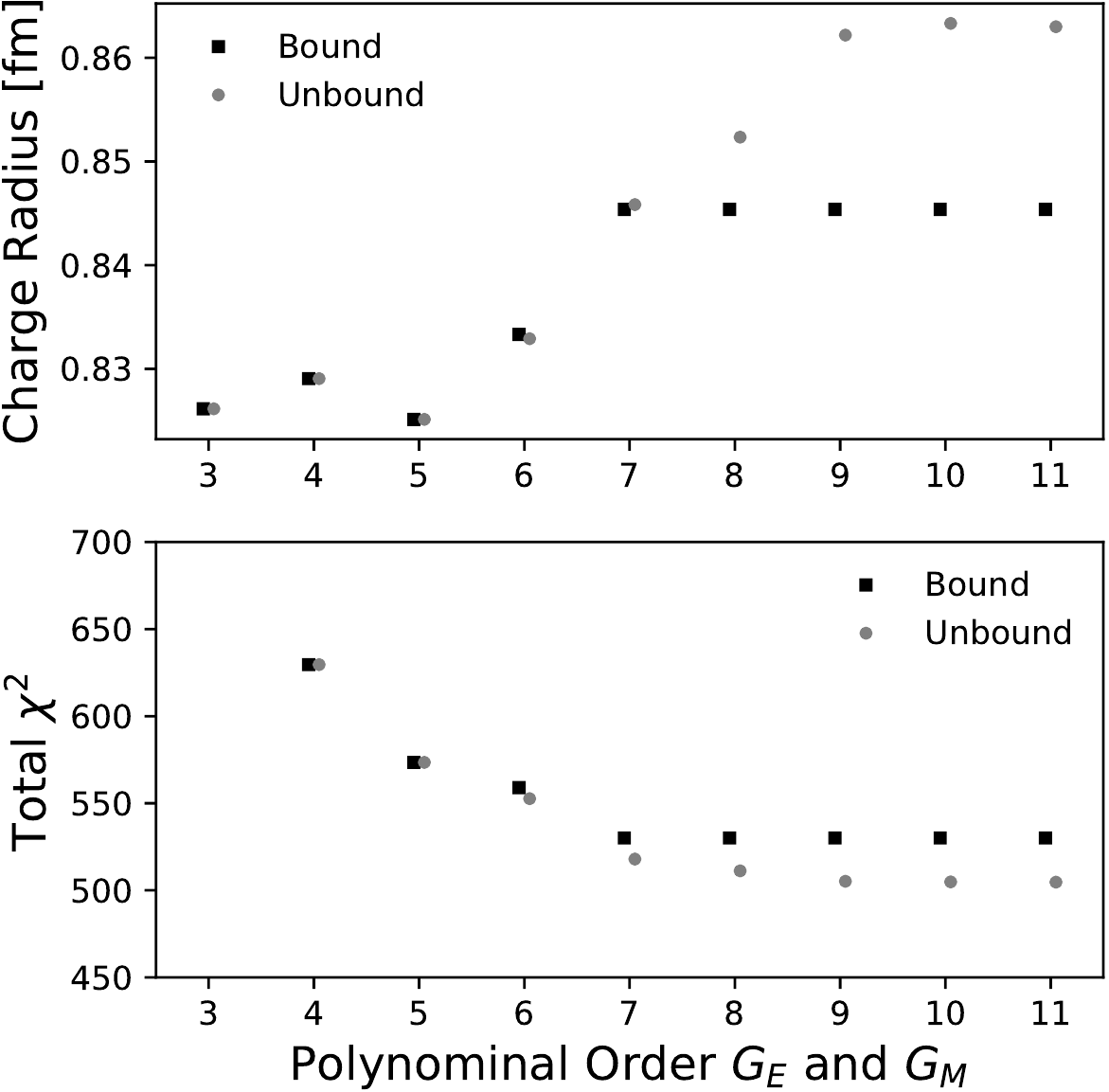}
\caption{Total $\chi^2$ vs. the number of fit parameters in $G_E$ and $G_M$ using Eq.~\ref{eq:polynomial} and~\ref{eq:polynomial2} for both the unbound and bound polynomial regressions of the 658 points of the rebinned Mainz dataset~\cite{Lee:2015jqa}.  
For these fits, by AIC and BIC, $7^{th}$ order is the most appropriate for the bound regression while for the unbound 9$^{th}$ order is the most appropriate.}
\label{OtherChi2}
\end{figure}

For a fixed number of fit parameters, the unbound regressions presented in this work will always have a total $\chi^2$ equal to or lower than a bound regression as shown in Fig.~\ref{TheChi2} and Fig.~\ref{OtherChi2} which plot the total $\chi^2$ and charge radius given by the unbound and bound regressions of the 1422 Mainz cross section points and the 658 rebinned cross sections respectively. Since adding parameters will always either decrease or keep total $\chi^2$ the same, $\chi^2$ by itself is not a valid model selection criterion. More appropriate model selection techniques include using an F-Test for nested models that are linear in terms~\cite{Higinbotham:2015rja,Sirca:2016} or model selection methods like the Akaike Information Criterion (AIC)~\cite{Akaike:1974} or the Bayesian Information Criterion (BIC)~\cite{Schwarz:1978} which can be used with non-nested non-linear models (see Ref.~\cite{Higinbotham:2018jfh} for more details). 

Since the regressions herein are non-linear, we use the AIC and BIC to determine the most appropriate number of parameters for these regressions. These statistical criteria, along with the frequently quoted 
$\chi^2$ per degree of freedom  $\chi^2/{\mathrm{df}}$~\cite{Rene:2010}, are defined as follows:
\begin{align}
\chi^2         & = \sum_{n=1}^{N}((data_i - model_i) / \sigma_i)^2, \\
\chi^2/{\mathrm{df}}& = \chi^2/ (N - N_{\mathrm{var}}), \\
\mathrm{AIC}            & = N \log(\chi^2/N) + 2 N_{\mathrm{var}}, \\
\mathrm{BIC}            & = N \log(\chi^2/N) + \log(N) N_{\mathrm{var}}, 
\end{align}
where $N$ is the number of data points, $data_i$ and $\sigma_i$ are the measured values and their estimated uncertainties respectively, $model_i$ is the model value, and $N_{\mathrm{var}}$ is the number of model parameters. Starting from lowest order fits, at first the value of these criteria will decrease as a parameter is added indicating an underfitting of the data while eventually the criteria will start to increase as parameters are added indicating an overfitting of the data; thus, with these criteria the model with the lowest AIC or BIC value should be selected as most appropriate.

For the 1422 original Mainz data points, we find with both AIC and BIC the most appropriate of the bound fits is the 7$^{th}$ order fit with a $\chi^2/{\mathrm{df}}$ of 1.21, while for the unbound fits the 10$^{th}$ order fit with a $\chi^2/{\mathrm{df}}$ of 1.14 is most appropriate as previously found~\cite{Bernauer:2010wm,Bernauer:2013tpr}. For the 658 rebinned Mainz data points, we find with both AIC and BIC the most appropriate of the bound fits is the 7$^{th}$ order fit with a $\chi^2/{\mathrm{df}}$ of 0.865, while for the unbound fits the 9$^{th}$ order fit with a $\chi^2/{\mathrm{df}}$ of 0.830 is most appropriate. Ideally a criteria would have been defined prior to the analysis of the data, since we are doing a re-analysis, we are presenting two of the most common techniques so one can judge how even the criteria can effect the outcome of the analysis.   
Further details about model selection techniques can be found in~\cite{Ernst:2012}. 

One should also keep in mind whether one is trying to do a descriptive fit of the data or, by adding physical constraints, building a predictive or explanatory model of the data~\cite{shmueli2010}.   
One must also keep in mind that none of these model selection techniques will prevent the use of completely inappropriate functions nor do they ensure that the best type of function has been selected. 
As noted in Ref.~\cite{Anscombe:1973,Rene:2010}, it is essential to plot the fit functions and residuals to ensure a 
reasonable regression as $\chi^2$ minimization alone is insufficient as 
illustrated in Appendix~\ref{appendix-quartet}.

As the changes we have presented in these four fits are larger than the statistical parameter uncertainties, 
we have limited ourselves to a discussion of the shifts of the mean values of the points. For non-linear 
regressions such as these, statistical bootstrapping which makes use of sampling with replacement can 
be used to find the statistical parameter uncertainties~\cite{Recipes:2007}.    

\section{Results}

\begin{figure}[htb]
\includegraphics[width=\columnwidth]{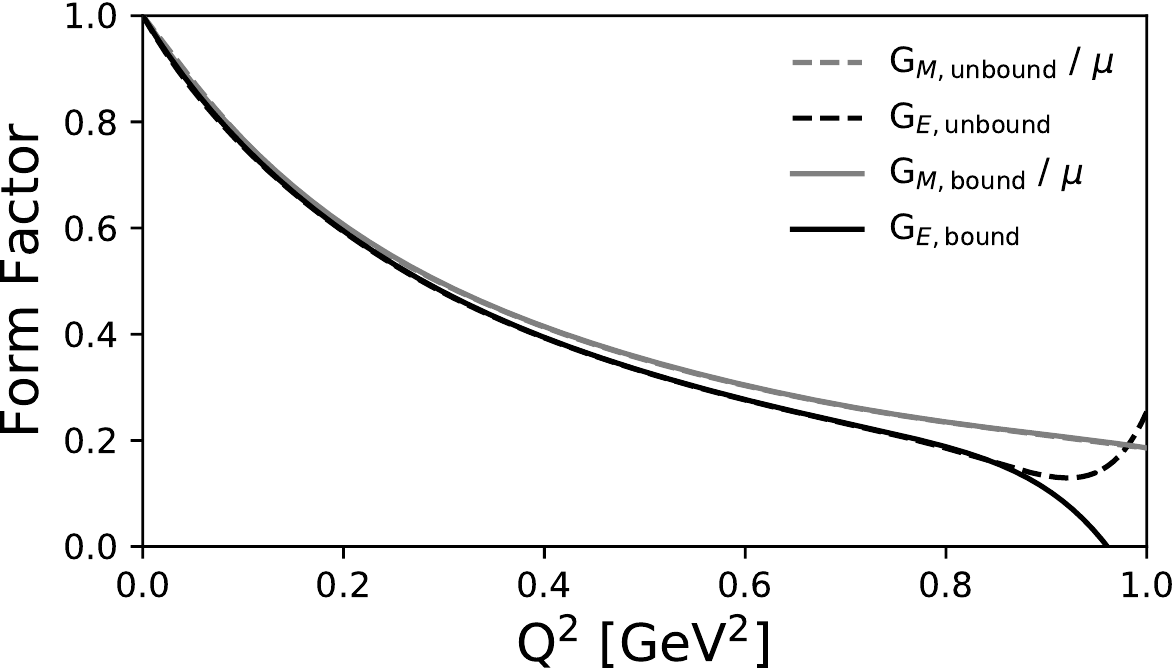}
\caption{The electromagnetic form factors vs. $Q^2$ from the unbound and bound polynomial regressions of the 1422 Mainz cross sections~\cite{Bernauer:2013tpr}. For these kinematics, as the $Q^2$ gets large, the cross sections are dominated by $G_M$ and the $G_E$ form factor becomes unconstrained, so the divergence of $G_E$ at high $Q^2$ is to be expected from a high-order polynomial.}
\label{fig:form}
\end{figure}

\begin{figure}[htb]
\includegraphics[width=\columnwidth]{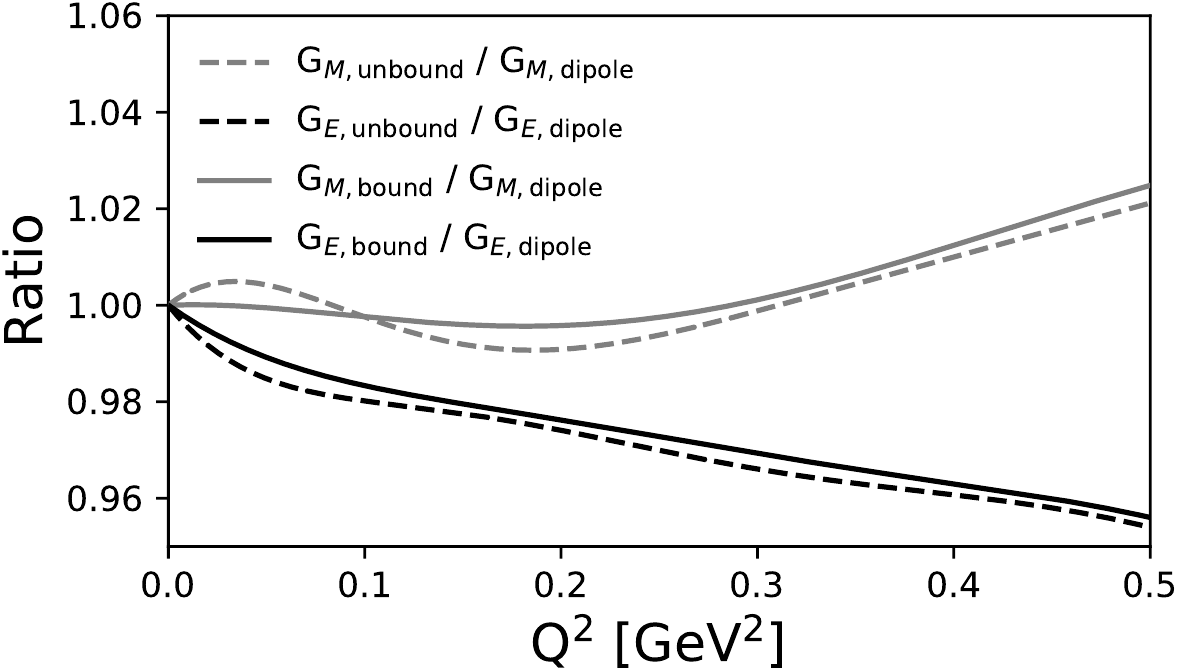}
\caption{The ratio of the extracted electromagnetic form factors to standard dipole vs. $Q^2$ for the 1422 Mainz cross sections. The oscillations in the unbound magnetic form factor, and to a lesser extent electric form factor, go away once the terms of the fit functions are forced to alternate sign.   The smooth magnetic form factor also results in a smaller extracted proton charge radius (0.854~fm vs. 0.882~fm).}
\label{fig:formdivdip}
\end{figure}

\noindent In Fig.~\ref{fig:form} and Fig.~\ref{fig:formdivdip}, we show the individual electric and magnetic form factors obtained from the unbound and bound regressions for the 1422 Mainz cross section points. In Fig.~\ref{fig:form} we see that $G_M$ for both the unbound and bound fits remains well behaved at high $Q^2$. However, for $G_E$ both the unbound and bound fits begin to diverge at high $Q^2$. This is due to the dominance of the magnetic form factor in the cross section at high $Q^2$ which leads the electric form factor to become unconstrained in this region. Due to using a high-order polynomial regression, along with the unconstrained nature of $G_E$ in the high $Q^2$ region, the divergence of $G_E$ is to be expected.

In Fig.~\ref{fig:formdivdip} the ratios of the electromagnetic form factors to the standard dipole are shown. This ratio reveals that the unbound $G_E$ ratio has a small oscillation and the unbound $G_M$ ratio has a large oscillation. Whereas, the bound $G_E$ ratio has mostly removed the oscillation and the bound $G_M$ ratio has almost no oscillation. The unbound ratios are descriptive models of the scattering data without any physics considerations, but the bound ratios are more akin to predictive models as the terms alternate sign as one would expect from chiral effective field theory. By adding this one physical constraint the oscillations are removed as the model becomes more predictive. 

Though it is beyond the range of the data used in the regression, the results of regressions like these are frequently used to extract the charge radius of the proton by using Eq.~\ref{eq:radius} to relate the fit function to the charge radius of the proton. Ideally, these extractions would use a predictive model to fit the data as finding the charge radius requires extrapolating beyond the range of the experimental data. For the case of a polynomial regression, this is simply:

\begin{equation}
r_p = ( -6 a^E_1 )^{1/2}\text{.}
\end{equation}

\noindent Using Eq.~\ref{eq:radius} one finds a charge radius of 0.882~fm from the unbound regression and 0.854~fm from the bound regression of the original Mainz cross section points (Table~\ref{poly:madness}), and a charge radius of 0.863~fm from the unbound regression and 0.845~fm from the bound regression of the rebinned Mainz data (Table~\ref{table:ff658}). These results show that even just rebinning the data can shift the result of a high-order polynomial regression significantly.

In the end, the radii extracted from the more descriptive unbound regressions are closer to the CODATA-2014 value, while the radii extracted from the more physically justifiable and predictive bound regressions are closer to the muonic results as well as the most recent atomic result~\cite{Bezginov:2019}. With freedom to make analytic choices that so strongly affect the results, there is the potential for unconscious confirmation bias, and for researchers to select and report the regressions that confirm their expectations~\cite{bias,false}.

%
%
%
%

\section{Conclusions}

We have shown that small changes in analytic functions and binning choices applied to a complex non-linear regression can result in significantly different results. In particular, using the Mainz dataset of elastic cross section points to extract a proton charge radius, we have shown results consistent with the CODATA-2014 value when using high-order unbounded polynomial fits and values close to the muonic results when using bounded polynomial regressions. Thus, by simply trying different functions, limits, and bounds, one can easily extrapolate to different results which can lead to confirmation bias and/or inappropriate rejection of certain results. Enrico Fermi noted that these types of problems should be addressed using either a firm mathematical basis or a physical model~\cite{Dyson:2004}. Ideally, regression models should be carefully developed prior to obtaining experimental data, as was done by the PRad collaboration~\cite{Yan:2018bez,Xiong:2019umf}; otherwise, one must be exceedingly careful to avoid confirmation bias though the rigorous use of model selection techniques~\cite{bias,false}.

Thus, while one can argue that the bounded non-linear regression is the more physical function, it would be more appropriate to approach the analysis such that the analytic choices do not so strongly affect the results. To do this, one can either fit only lower $Q^2$ data where fewer free parameters are required~\cite{Hand:1963zz,Murphy:1974zz,Borkowski:1975ume,Simon:1980hu,
Higinbotham:2015rja,Griffioen:2015hta,Yan:2018bez,Hagelstein:2018zrz,Higinbotham:2018jfh,Hayward:2018qij} and the results are not sensitive to the magnetic form factor, as shown explicitly in Ref.~\cite{Higinbotham:2018jfh}; or, as Fermi preferred, use a physical model, such as that of Bernard~{\it{et al.}}~\cite{Bernard:1998gv} or Alarc\'{o}n and Weiss~\cite{Alarcon:2018irp} to constrain the fits~\cite{Hohler:1976ax,Lorenz:2014yda,Horbatsch:2016ilr,Alarcon:2018zbz}. There are also the physically motivated functions such as rational functions~\cite{Higinbotham:2015rja,Kelly:2004hm}, continued fractions~\cite{Griffioen:2015hta,Sick:2003gm,Lorenz:2012tm}, or the z-expansion fits~\cite{Hill:2010yb,Lee:2015jqa} though these still require model selection techniques to determine the appropriate number of regression parameters. We hope to have illustrated that by using extremely complex non-linear regressions and deep searches, one can find nearly any radius in a wide range of radii from a single dataset~\cite{Higinbotham-Code-2019}. To quote Nobel laureate Ronald H. Coase, ``if you torture the data long enough, it will confess.''   

\acknowledgments
Many thanks to Franziska Hagelstein and Vladimir Pascalutsa for their questions about normalization parameters that prompted this work. Thanks to Dave Meekins for his many useful comments and discussions. Thanks to Nigel Tufnel for suggesting going to eleven. And thanks to Marcy Stutzman and Simon \v{S}irca for their editorial comments. This work was supported by the U.S. Department of Energy contract DE-AC05-06OR23177 under which Jefferson Science Associates operates the Thomas Jefferson National Accelerator Facility.

\appendix

\section{Robust Regressions}
\label{sec:robust}

Ordinary least squares regression (OLSR) is one of the most commonly employed techniques used to fit a given model and its parameters to a dataset. However, OLSR is commonly misunderstood and misapplied by researchers. For OLSR, fit parameters are determined by minimizing the square of the differences between real-world data and model predictions. This is known as a $\chi^2$ minimization. Eq.~\ref{eq:chi2} shows this minimization for $N$ data points with $M$ fit parameters.

\begin{equation}\label{eq:chi2}
	\chi^2 \equiv \sum^N_{i=1} \left( \frac{y_i - y\left( x_i | a_1, a_2, ..., a_M \right)}{\sigma_i} \right)^2
\end{equation}

\noindent Here $y_i$ are the measured data values, $y\left( x_i | a_1, a_2, ..., a_M \right)$ are the values given by the model with fit parameters $a_1$ to $a_M$ when evaluated at the $x_i$ of the measured data, and $\sigma_i$ are the uncertainties on each measured data point.

While OLSR via $\chi^2$ minimization is often a useful initial method for checking the `goodness' of a fit, it can fail if the dataset being fit does not meet certain conditions. OLSR is based on the core assumptions that the errors are random variables that are normally distributed, the errors are uncorrelated to each other, and the errors are homoscedastic, which is to say they have the same variance. Unfortunately, these assumptions often do not hold true in the case of real-world data. When OLSR's assumptions are not met, such as when the dataset has significant outliers, OLSR is not sufficient for fitting the data and can yield misleading results. Even a singular outlier can skew the results of an OLSR pulling the fit away from the data's true behavior~\cite{Recipes:2007,Draper:robust}. To avoid these pitfalls, robust methods such as robust least squares regression (RLSR) should be used instead of OLSR techniques.

To avoid outliers having too much influence over a fit, we desire a method by which outliers can be identified and then re-weighted such that they do not skew the overall fit. The least squares minimization found in Eq.~\ref{eq:chi2} can be generalized to Eq.~\ref{eq:rlsq} by introducing the function $\rho(z)$~\cite{triggs:2000}. OLSR is then simply the case where $\rho(z)=z$. Many functions can be used for $\rho(z)$ to introduce robustness, but for the following examples the `soft loss' (softl1) function given in Eq.~\ref{eq:softl1} was selected and implemented using the Python package SciPy~\cite{Oliphant:2001,Oliphant:2007,triggs:2000}. 

\begin{align}\label{eq:rlsq}
	\chi^2 \equiv \sum^N_{i=1} \rho_i\left( z \right) \; \text{and} \; z = \left( \frac{y_i - y\left( x_i | a_1, a_2, ..., a_M \right)}{\sigma_i} \right)^2
\end{align}

\begin{equation}\label{eq:softl1}
	\rho \left(z\right) = 2 \left(\sqrt{1 + z} - 1\right)
\end{equation}

With soft loss, as a $z_i$ gets larger, the magnitude of $\rho_i(z)$ is increasingly reduced with respect to OLSR. A RLSR with soft loss essentially re-weights the outliers of a dataset, decreasing their influence when fitting.  Note that if a dataset meets all of the above assumptions inherent to OLSR (i.e. errors are normally distributed, uncorrelated, and have the same variance) then OLSR and RLSR techniques should both produce the same fit results since the dataset, by definition, does not contain excessive outliers.

\begin{figure}[htb]
\includegraphics[width=\columnwidth]{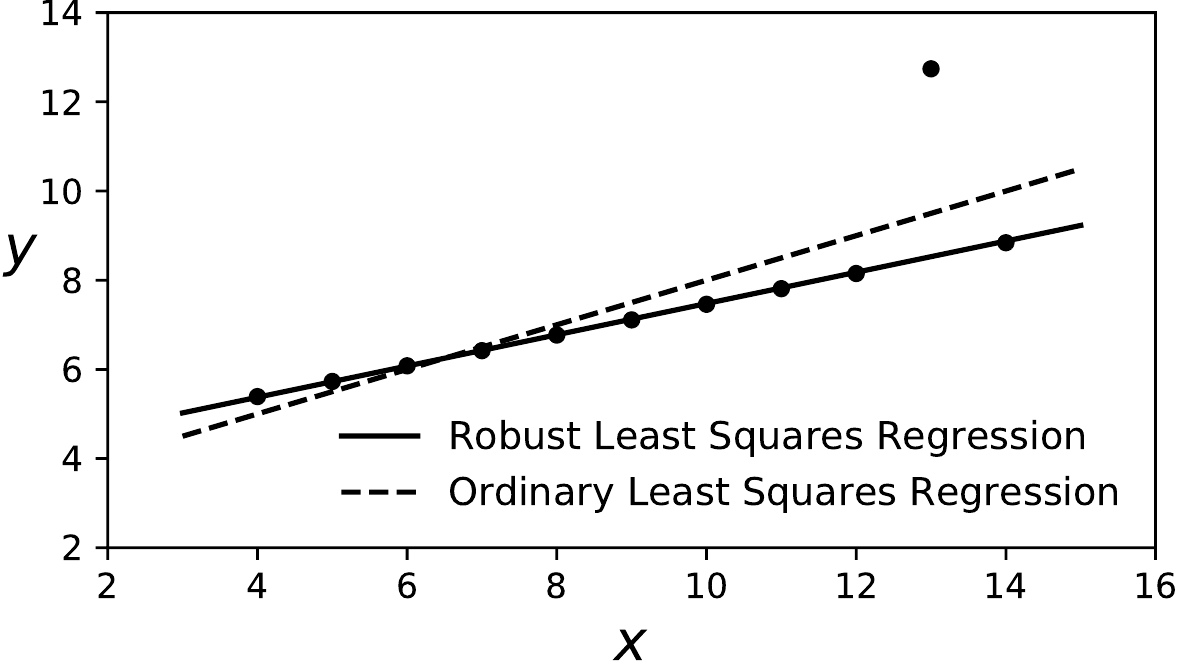}
\caption{This example data, reproduced from a classic statistics paper~\cite{Anscombe:1973}, shows how an ordinary least squares regression, OLSR, is easily pulled away from the true trend of the data while a robust least squares regression, RLSR, is only weakly affected by the outlying data point.} 
\label{fig:robust}
\end{figure}

A simple example reproduced from a classic statistics paper~\cite{Anscombe:1973} is shown in Fig.~\ref{fig:robust}. The dataset has a single clear outlier which pulls the fit considerably away from the bulk of the data when using OLSR. However, when RLSR with soft loss is used to fit the data the influence of the outlier is greatly reduced, and the fit returns to the bulk of the data yielding superior results.  

\begin{figure}[htb]
\includegraphics[width=\columnwidth]{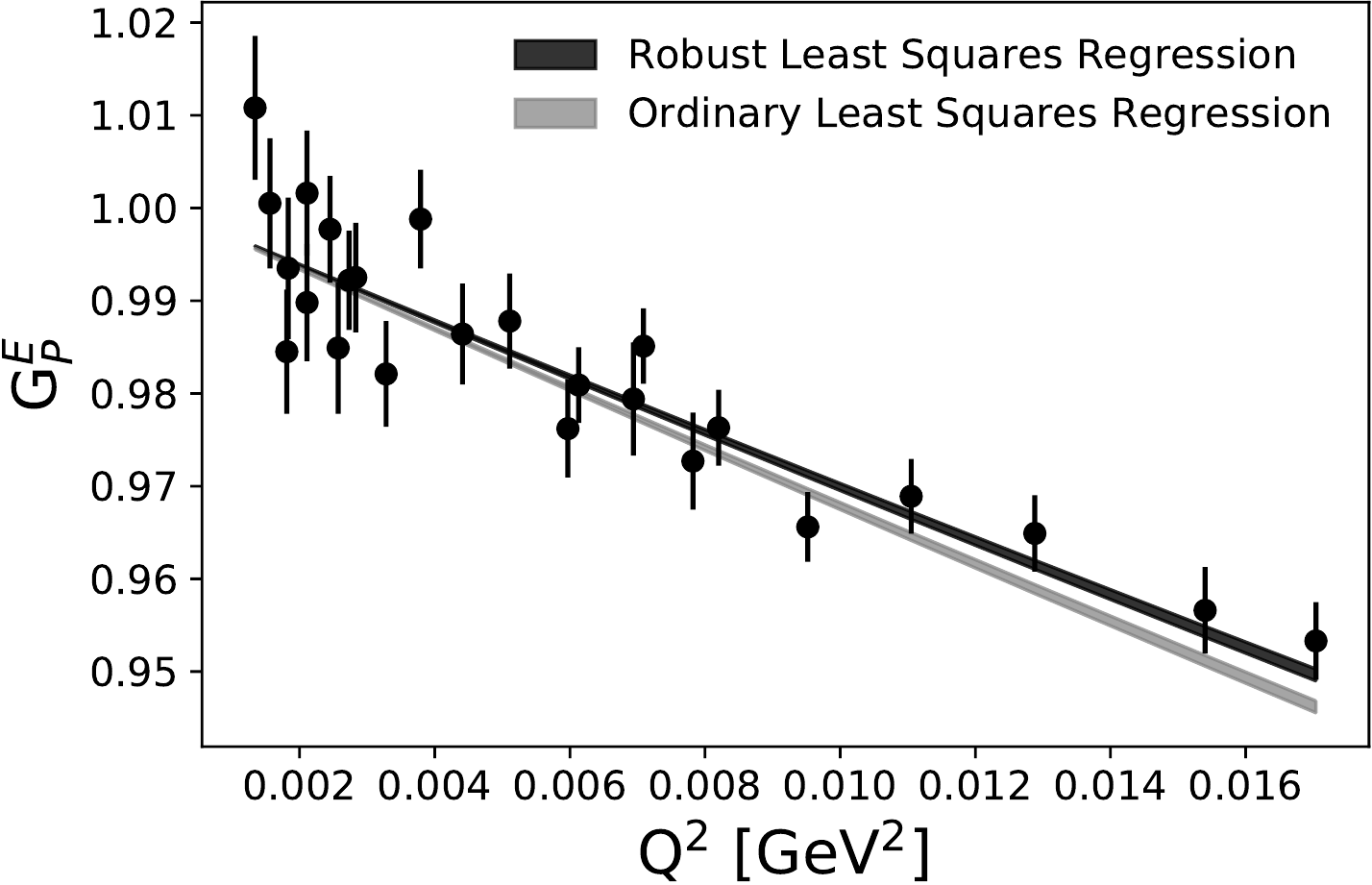}
\caption{Proton electric form factor data taken from the Initial State Radiation dataset found in the supplemental material of Ref.~\cite{Mihovilovic:2019jiz}. The uncertainties are calculated by summing the listed statistical uncertainties with the systematic uncorrelated uncertainties in quadrature. The theoretical model used for the regressions is the model of Alarc\'{o}n and Weiss~\cite{Alarcon:2018irp}, with only one free parameter. These regressions give a proton radius of 0.874~fm for the OLSR and 0.844~fm for the RLSR with soft loss.} 
\label{fig:miha}
\end{figure}

For a real-world example using RLSR we can study the full Initial State Radiation (ISR) dataset found in the supplemental material of Ref.~\cite{Mihovilovic:2019jiz,Mihovilovic:2016rkr}. Fig.~\ref{fig:miha} shows the results of two regressions of the ISR proton electric form factor data using the theory model of Alarc\'{o}n and Weiss~\cite{Alarcon:2018irp}, with the proton radius as its single free parameter. The light gray curve uses an OLSR to fit the dataset and the dark curve uses a RLSR with soft loss. 

There is a clear separation between the results of the two regressions, with the RLSR fitting the higher $Q^2$ data better. The OLSR finds a proton radius of 0.874 fm, and the RLSR finds a significantly smaller radius of 0.844 fm. Again, the purely analytic choice of the regression type significantly influences the fit results. Further, the fact that the OLSR and the RLSR fit results differ significantly is evidence that there are outliers in the dataset that are not following a normal distribution, otherwise the OLSR and RLSR fits would have better agreement.

\section{Anscombe's Quartet}
\label{appendix-quartet}
With the power of modern computing, one can be tempted to blindly assume the results of the regression are correct if the $\chi^2/{\mathrm{df}}$ is close to unity; but this can be extremely misleading~\cite{Sirca:2012}.  To illustrate
this point we use the Anscombe quartet~\cite{Anscombe:1973}, though as nuclear physicists tend to use $\chi^2/{\mathrm{df}}$ instead of R$^2$. We have taken the 1973 example problem and added uncertainties to the points as shown in Table~\ref{quartet}.

\begin{table}[htb]
\caption{Four datasets of (x$_i$,y$_i$,dy$_i$) values adapted from~\cite{Anscombe:1973}.} 
\begin{tabular}{ccccccc}
  x$_{1,2,3}$      & y$_{1}$     & y$_{2}$     & y$_{3}$     & x$_{4}$    &  y$_{4}$    & dy$_{1,2,3,4}$      \\  \hline
          10.0 & 8.04 & 9.14 & 7.46 & 8.0 & 6.58 & 1.235 \\
           8.0 & 6.95 & 8.14 & 6.77 & 8.0 & 5.76 & 1.235 \\
          13.0 & 7.58 & 8.74 &12.74 & 8.0 & 7.71 & 1.235 \\
           9.0 & 8.81 & 8.77 & 7.11 & 8.0 & 8.84 & 1.235 \\
          11.0 & 8.33 & 9.26 & 7.81 & 8.0 & 8.47 & 1.235 \\
          14.0 & 9.96 & 8.10 & 8.84 & 8.0 & 7.04 & 1.235 \\
           6.0 & 7.24 & 6.13 & 6.08 & 8.0 & 5.25 & 1.235 \\
           4.0 & 4.26 & 3.10 & 5.39 &19.0 &12.50 & 1.235 \\
          12.0 &10.84 & 9.13 & 8.15 & 8.0 & 5.56 & 1.235 \\
          7.0 & 4.82 & 7.26 & 6.42 & 8.0 & 7.91 & 1.235 \\
          5.0 & 5.68 & 4.74 & 5.73 & 8.0 & 6.89 & 1.235 \\ \hline
\end{tabular}
\label{quartet}
\end{table}

\begin{figure}[htb]
\includegraphics[width=\columnwidth]{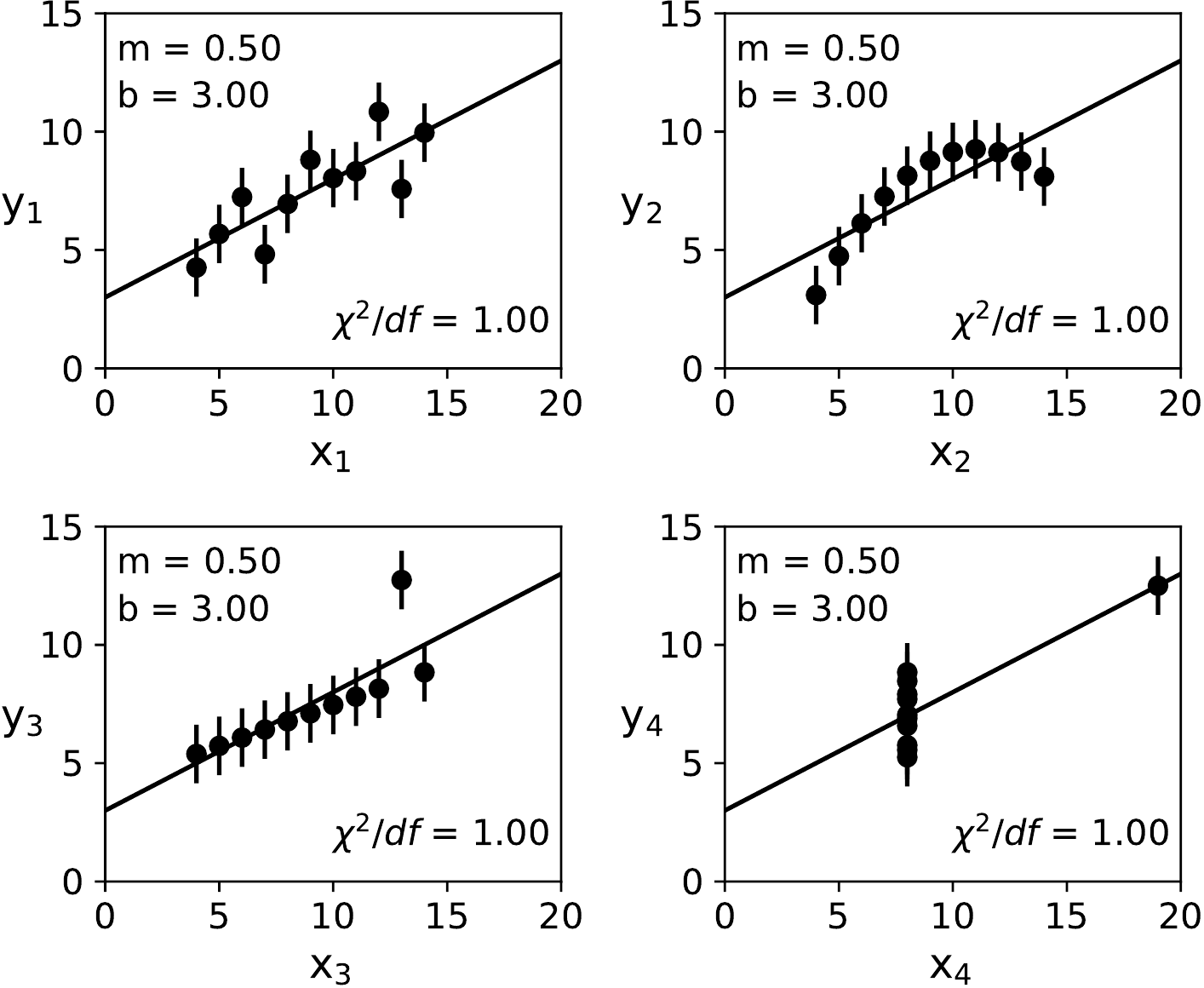}
\caption{Graphs of the four sets of data from Table~\ref{quartet}.   When fit with a linear function, $y=mx+b$, all
four sets of data give the same slope, same intercept and same $\chi^2/df$ of one; yet clearly these four sets of data are not the same.}
\label{SameChi2}
\end{figure}

When fit with a linear function, these four sets of (x,y,dy) values give 
the same statistical quantities to three significant figures: 
mean, variance, $\chi^2$, $\chi^2/{\mathrm{df}}$, etc.
So if one fails to make graphical checks, one can be completely fooled into thinking
the fits are all equally good; but by simply graphing (see Fig.~\ref{SameChi2})
one can see that only the first set
of data is distributed in an ideal way around the fit function.

Dataset two clearly has a curved residual yet has exactly the same mean, errors, and $\chi^2$ as the first fit. This suggests that the fitter should likely add a quadratic term to the regression as well as check that the uncertainties have been correctly reported.

Dataset three illustrates the effect of an outlier on the regression. Of course, the scientist doesn't simply throw out an outlier. Instead one should report on the outlier's influence on the result. For example, in dataset three it would be worth noting that if the outlier is removed, the data exactly follow a line of $y = 4 + 0.346 x$ and that that measurement should be repeated.

Dataset four looks unsatisfactory, since all the information about the slope comes from one observation. This is very different from dataset one where any one point can be removed and one will obtain nearly the same result. Thus, for dataset four it should be pointed out that a single observation plays a critical role in the result.


\begin{thebibliography}{59}%
\makeatletter
\providecommand \@ifxundefined [1]{%
 \@ifx{#1\undefined}
}%
\providecommand \@ifnum [1]{%
 \ifnum #1\expandafter \@firstoftwo
 \else \expandafter \@secondoftwo
 \fi
}%
\providecommand \@ifx [1]{%
 \ifx #1\expandafter \@firstoftwo
 \else \expandafter \@secondoftwo
 \fi
}%
\providecommand \natexlab [1]{#1}%
\providecommand \enquote  [1]{``#1''}%
\providecommand \bibnamefont  [1]{#1}%
\providecommand \bibfnamefont [1]{#1}%
\providecommand \citenamefont [1]{#1}%
\providecommand \href@noop [0]{\@secondoftwo}%
\providecommand \href [0]{\begingroup \@sanitize@url \@href}%
\providecommand \@href[1]{\@@startlink{#1}\@@href}%
\providecommand \@@href[1]{\endgroup#1\@@endlink}%
\providecommand \@sanitize@url [0]{\catcode `\\12\catcode `\$12\catcode
  `\&12\catcode `\#12\catcode `\^12\catcode `\_12\catcode `\%12\relax}%
\providecommand \@@startlink[1]{}%
\providecommand \@@endlink[0]{}%
\providecommand \url  [0]{\begingroup\@sanitize@url \@url }%
\providecommand \@url [1]{\endgroup\@href {#1}{\urlprefix }}%
\providecommand \urlprefix  [0]{URL }%
\providecommand \Eprint [0]{\href }%
\providecommand \doibase [0]{http://dx.doi.org/}%
\providecommand \selectlanguage [0]{\@gobble}%
\providecommand \bibinfo  [0]{\@secondoftwo}%
\providecommand \bibfield  [0]{\@secondoftwo}%
\providecommand \translation [1]{[#1]}%
\providecommand \BibitemOpen [0]{}%
\providecommand \bibitemStop [0]{}%
\providecommand \bibitemNoStop [0]{.\EOS\space}%
\providecommand \EOS [0]{\spacefactor3000\relax}%
\providecommand \BibitemShut  [1]{\csname bibitem#1\endcsname}%
\let\auto@bib@innerbib\@empty
\bibitem [{\citenamefont {Silberzahn}\ \emph {et~al.}(2018)\citenamefont
  {Silberzahn} \emph {et~al.}}]{Silberzahn:2018}%
  \BibitemOpen
  \bibfield  {author} {\bibinfo {author} {\bibfnamefont {R.}~\bibnamefont
  {Silberzahn}} \emph {et~al.},\ }\href {\doibase 10.1177/2515245917747646}
  {\bibfield  {journal} {\bibinfo  {journal} {Advances in Methods and Practices
  in Psychological Science}\ }\textbf {\bibinfo {volume} {1}},\ \bibinfo
  {pages} {337} (\bibinfo {year} {2018})}\BibitemShut {NoStop}%
\bibitem [{\citenamefont {Shmueli}(2010)}]{shmueli2010}%
  \BibitemOpen
  \bibfield  {author} {\bibinfo {author} {\bibfnamefont {G.}~\bibnamefont
  {Shmueli}},\ }\href {\doibase 10.1214/10-STS330} {\bibfield  {journal}
  {\bibinfo  {journal} {Statist. Sci.}\ }\textbf {\bibinfo {volume} {25}},\
  \bibinfo {pages} {289} (\bibinfo {year} {2010})}\BibitemShut {NoStop}%
\bibitem [{\citenamefont {Pohl}\ \emph {et~al.}(2010)\citenamefont {Pohl} \emph
  {et~al.}}]{Pohl:2010zza}%
  \BibitemOpen
  \bibfield  {author} {\bibinfo {author} {\bibfnamefont {R.}~\bibnamefont
  {Pohl}} \emph {et~al.},\ }\href {\doibase 10.1038/nature09250} {\bibfield
  {journal} {\bibinfo  {journal} {Nature}\ }\textbf {\bibinfo {volume} {466}},\
  \bibinfo {pages} {213} (\bibinfo {year} {2010})}\BibitemShut {NoStop}%
\bibitem [{\citenamefont {Antognini}\ \emph {et~al.}(2013)\citenamefont
  {Antognini} \emph {et~al.}}]{Antognini:1900ns}%
  \BibitemOpen
  \bibfield  {author} {\bibinfo {author} {\bibfnamefont {A.}~\bibnamefont
  {Antognini}} \emph {et~al.},\ }\href {\doibase 10.1126/science.1230016}
  {\bibfield  {journal} {\bibinfo  {journal} {Science}\ }\textbf {\bibinfo
  {volume} {339}},\ \bibinfo {pages} {417} (\bibinfo {year}
  {2013})}\BibitemShut {NoStop}%
\bibitem [{\citenamefont {Mohr}\ \emph {et~al.}(2016)\citenamefont {Mohr},
  \citenamefont {Newell},\ and\ \citenamefont {Taylor}}]{Mohr:2015ccw}%
  \BibitemOpen
  \bibfield  {author} {\bibinfo {author} {\bibfnamefont {P.~J.}\ \bibnamefont
  {Mohr}}, \bibinfo {author} {\bibfnamefont {D.~B.}\ \bibnamefont {Newell}}, \
  and\ \bibinfo {author} {\bibfnamefont {B.~N.}\ \bibnamefont {Taylor}},\
  }\href {\doibase 10.1103/RevModPhys.88.035009} {\bibfield  {journal}
  {\bibinfo  {journal} {Rev. Mod. Phys.}\ }\textbf {\bibinfo {volume} {88}},\
  \bibinfo {pages} {035009} (\bibinfo {year} {2016})},\ \Eprint
  {http://arxiv.org/abs/1507.07956} {arXiv:1507.07956 [physics.atom-ph]}
  \BibitemShut {NoStop}%
\bibitem [{\citenamefont {Pohl}\ \emph {et~al.}(2013)\citenamefont {Pohl},
  \citenamefont {Gilman}, \citenamefont {Miller},\ and\ \citenamefont
  {Pachucki}}]{Pohl:2013yb}%
  \BibitemOpen
  \bibfield  {author} {\bibinfo {author} {\bibfnamefont {R.}~\bibnamefont
  {Pohl}}, \bibinfo {author} {\bibfnamefont {R.}~\bibnamefont {Gilman}},
  \bibinfo {author} {\bibfnamefont {G.~A.}\ \bibnamefont {Miller}}, \ and\
  \bibinfo {author} {\bibfnamefont {K.}~\bibnamefont {Pachucki}},\ }\href
  {\doibase 10.1146/annurev-nucl-102212-170627} {\bibfield  {journal} {\bibinfo
   {journal} {Ann. Rev. Nucl. Part. Sci.}\ }\textbf {\bibinfo {volume} {63}},\
  \bibinfo {pages} {175} (\bibinfo {year} {2013})},\ \Eprint
  {http://arxiv.org/abs/1301.0905} {arXiv:1301.0905 [physics.atom-ph]}
  \BibitemShut {NoStop}%
\bibitem [{\citenamefont {Carlson}(2015)}]{Carlson:2015jba}%
  \BibitemOpen
  \bibfield  {author} {\bibinfo {author} {\bibfnamefont {C.~E.}\ \bibnamefont
  {Carlson}},\ }\href {\doibase 10.1016/j.ppnp.2015.01.002} {\bibfield
  {journal} {\bibinfo  {journal} {Prog. Part. Nucl. Phys.}\ }\textbf {\bibinfo
  {volume} {82}},\ \bibinfo {pages} {59} (\bibinfo {year} {2015})},\ \Eprint
  {http://arxiv.org/abs/1502.05314} {arXiv:1502.05314 [hep-ph]} \BibitemShut
  {NoStop}%
\bibitem [{\citenamefont {Miller}(2018)}]{Miller:2018zfu}%
  \BibitemOpen
  \bibfield  {author} {\bibinfo {author} {\bibfnamefont {G.~A.}\ \bibnamefont
  {Miller}},\ }in\ \href@noop {} {\emph {\bibinfo {booktitle} {{13th Conference
  on the Intersections of Particle and Nuclear Physics (CIPANP 2018) Palm
  Springs, California, USA, May 29-June 3, 2018}}}}\ (\bibinfo {year} {2018})\
  \Eprint {http://arxiv.org/abs/1809.09635} {arXiv:1809.09635
  [physics.atom-ph]} \BibitemShut {NoStop}%
\bibitem [{\citenamefont {Miller}(2019)}]{Miller:2018ybm}%
  \BibitemOpen
  \bibfield  {author} {\bibinfo {author} {\bibfnamefont {G.~A.}\ \bibnamefont
  {Miller}},\ }\href {\doibase 10.1103/PhysRevC.99.035202} {\bibfield
  {journal} {\bibinfo  {journal} {Phys. Rev.}\ }\textbf {\bibinfo {volume}
  {C99}},\ \bibinfo {pages} {035202} (\bibinfo {year} {2019})},\ \Eprint
  {http://arxiv.org/abs/1812.02714} {arXiv:1812.02714 [nucl-th]} \BibitemShut
  {NoStop}%
\bibitem [{\citenamefont {Horbatsch}\ and\ \citenamefont
  {Hessels}(2016)}]{Horbatsch:2015qda}%
  \BibitemOpen
  \bibfield  {author} {\bibinfo {author} {\bibfnamefont {M.}~\bibnamefont
  {Horbatsch}}\ and\ \bibinfo {author} {\bibfnamefont {E.~A.}\ \bibnamefont
  {Hessels}},\ }\href {\doibase 10.1103/PhysRevC.93.015204} {\bibfield
  {journal} {\bibinfo  {journal} {Phys. Rev.}\ }\textbf {\bibinfo {volume}
  {C93}},\ \bibinfo {pages} {015204} (\bibinfo {year} {2016})},\ \Eprint
  {http://arxiv.org/abs/1509.05644} {arXiv:1509.05644 [nucl-ex]} \BibitemShut
  {NoStop}%
\bibitem [{\citenamefont {Griffioen}\ \emph {et~al.}(2016)\citenamefont
  {Griffioen}, \citenamefont {Carlson},\ and\ \citenamefont
  {Maddox}}]{Griffioen:2015hta}%
  \BibitemOpen
  \bibfield  {author} {\bibinfo {author} {\bibfnamefont {K.}~\bibnamefont
  {Griffioen}}, \bibinfo {author} {\bibfnamefont {C.}~\bibnamefont {Carlson}},
  \ and\ \bibinfo {author} {\bibfnamefont {S.}~\bibnamefont {Maddox}},\ }\href
  {\doibase 10.1103/PhysRevC.93.065207} {\bibfield  {journal} {\bibinfo
  {journal} {Phys. Rev.}\ }\textbf {\bibinfo {volume} {C93}},\ \bibinfo {pages}
  {065207} (\bibinfo {year} {2016})},\ \Eprint
  {http://arxiv.org/abs/1509.06676} {arXiv:1509.06676 [nucl-ex]} \BibitemShut
  {NoStop}%
\bibitem [{\citenamefont {Higinbotham}\ \emph {et~al.}(2016)\citenamefont
  {Higinbotham}, \citenamefont {Kabir}, \citenamefont {Lin}, \citenamefont
  {Meekins}, \citenamefont {Norum},\ and\ \citenamefont
  {Sawatzky}}]{Higinbotham:2015rja}%
  \BibitemOpen
  \bibfield  {author} {\bibinfo {author} {\bibfnamefont {D.~W.}\ \bibnamefont
  {Higinbotham}}, \bibinfo {author} {\bibfnamefont {A.~A.}\ \bibnamefont
  {Kabir}}, \bibinfo {author} {\bibfnamefont {V.}~\bibnamefont {Lin}}, \bibinfo
  {author} {\bibfnamefont {D.}~\bibnamefont {Meekins}}, \bibinfo {author}
  {\bibfnamefont {B.}~\bibnamefont {Norum}}, \ and\ \bibinfo {author}
  {\bibfnamefont {B.}~\bibnamefont {Sawatzky}},\ }\href {\doibase
  10.1103/PhysRevC.93.055207} {\bibfield  {journal} {\bibinfo  {journal} {Phys.
  Rev.}\ }\textbf {\bibinfo {volume} {C93}},\ \bibinfo {pages} {055207}
  (\bibinfo {year} {2016})},\ \Eprint {http://arxiv.org/abs/1510.01293}
  {arXiv:1510.01293 [nucl-ex]} \BibitemShut {NoStop}%
\bibitem [{\citenamefont {Lee}\ \emph {et~al.}(2015)\citenamefont {Lee},
  \citenamefont {Arrington},\ and\ \citenamefont {Hill}}]{Lee:2015jqa}%
  \BibitemOpen
  \bibfield  {author} {\bibinfo {author} {\bibfnamefont {G.}~\bibnamefont
  {Lee}}, \bibinfo {author} {\bibfnamefont {J.~R.}\ \bibnamefont {Arrington}},
  \ and\ \bibinfo {author} {\bibfnamefont {R.~J.}\ \bibnamefont {Hill}},\
  }\href {\doibase 10.1103/PhysRevD.92.013013} {\bibfield  {journal} {\bibinfo
  {journal} {Phys. Rev.}\ }\textbf {\bibinfo {volume} {D92}},\ \bibinfo {pages}
  {013013} (\bibinfo {year} {2015})},\ \Eprint
  {http://arxiv.org/abs/1505.01489} {arXiv:1505.01489 [hep-ph]} \BibitemShut
  {NoStop}%
\bibitem [{\citenamefont {Graczyk}\ and\ \citenamefont
  {Juszczak}(2014)}]{Graczyk:2014lba}%
  \BibitemOpen
  \bibfield  {author} {\bibinfo {author} {\bibfnamefont {K.~M.}\ \bibnamefont
  {Graczyk}}\ and\ \bibinfo {author} {\bibfnamefont {C.}~\bibnamefont
  {Juszczak}},\ }\href {\doibase 10.1103/PhysRevC.90.054334} {\bibfield
  {journal} {\bibinfo  {journal} {Phys. Rev.}\ }\textbf {\bibinfo {volume}
  {C90}},\ \bibinfo {pages} {054334} (\bibinfo {year} {2014})},\ \Eprint
  {http://arxiv.org/abs/1408.0150} {arXiv:1408.0150 [hep-ph]} \BibitemShut
  {NoStop}%
\bibitem [{\citenamefont {Lorenz}\ and\ \citenamefont
  {Mei{\ss}ner}(2014)}]{Lorenz:2014vha}%
  \BibitemOpen
  \bibfield  {author} {\bibinfo {author} {\bibfnamefont {I.~T.}\ \bibnamefont
  {Lorenz}}\ and\ \bibinfo {author} {\bibfnamefont {{\relax Ulf}.-G.}\
  \bibnamefont {Mei{\ss}ner}},\ }\href {\doibase
  10.1016/j.physletb.2014.08.010} {\bibfield  {journal} {\bibinfo  {journal}
  {Phys. Lett.}\ }\textbf {\bibinfo {volume} {B737}},\ \bibinfo {pages} {57}
  (\bibinfo {year} {2014})},\ \Eprint {http://arxiv.org/abs/1406.2962}
  {arXiv:1406.2962 [hep-ph]} \BibitemShut {NoStop}%
\bibitem [{\citenamefont {Horbatsch}\ \emph {et~al.}(2017)\citenamefont
  {Horbatsch}, \citenamefont {Hessels},\ and\ \citenamefont
  {Pineda}}]{Horbatsch:2016ilr}%
  \BibitemOpen
  \bibfield  {author} {\bibinfo {author} {\bibfnamefont {M.}~\bibnamefont
  {Horbatsch}}, \bibinfo {author} {\bibfnamefont {E.~A.}\ \bibnamefont
  {Hessels}}, \ and\ \bibinfo {author} {\bibfnamefont {A.}~\bibnamefont
  {Pineda}},\ }\href {\doibase 10.1103/PhysRevC.95.035203} {\bibfield
  {journal} {\bibinfo  {journal} {Phys. Rev.}\ }\textbf {\bibinfo {volume}
  {C95}},\ \bibinfo {pages} {035203} (\bibinfo {year} {2017})},\ \Eprint
  {http://arxiv.org/abs/1610.09760} {arXiv:1610.09760 [nucl-th]} \BibitemShut
  {NoStop}%
\bibitem [{\citenamefont {Alarc\'{o}n}\ \emph {et~al.}(2019)\citenamefont
  {Alarc\'{o}n}, \citenamefont {Higinbotham}, \citenamefont {Weiss},\ and\
  \citenamefont {Ye}}]{Alarcon:2018zbz}%
  \BibitemOpen
  \bibfield  {author} {\bibinfo {author} {\bibfnamefont {J.~M.}\ \bibnamefont
  {Alarc\'{o}n}}, \bibinfo {author} {\bibfnamefont {D.~W.}\ \bibnamefont
  {Higinbotham}}, \bibinfo {author} {\bibfnamefont {C.}~\bibnamefont {Weiss}},
  \ and\ \bibinfo {author} {\bibfnamefont {Z.}~\bibnamefont {Ye}},\ }\href
  {\doibase 10.1103/PhysRevC.99.044303} {\bibfield  {journal} {\bibinfo
  {journal} {Phys. Rev.}\ }\textbf {\bibinfo {volume} {C99}},\ \bibinfo {pages}
  {044303} (\bibinfo {year} {2019})},\ \Eprint
  {http://arxiv.org/abs/1809.06373} {arXiv:1809.06373 [hep-ph]} \BibitemShut
  {NoStop}%
\bibitem [{\citenamefont {Alarcón}\ \emph {et~al.}(2020)\citenamefont
  {Alarcón}, \citenamefont {Higinbotham},\ and\ \citenamefont
  {Weiss}}]{Alarcon:2020kcz}%
  \BibitemOpen
  \bibfield  {author} {\bibinfo {author} {\bibfnamefont {J.}~\bibnamefont
  {Alarcón}}, \bibinfo {author} {\bibfnamefont {D.}~\bibnamefont
  {Higinbotham}}, \ and\ \bibinfo {author} {\bibfnamefont {C.}~\bibnamefont
  {Weiss}},\ }\href@noop {} {\  (\bibinfo {year} {2020})},\ \Eprint
  {http://arxiv.org/abs/2002.05167} {arXiv:2002.05167 [hep-ph]} \BibitemShut
  {NoStop}%
\bibitem [{\citenamefont {Zhou}\ \emph {et~al.}(2019)\citenamefont {Zhou},
  \citenamefont {Giuliani}, \citenamefont {Piekarewicz}, \citenamefont
  {Bhattacharya},\ and\ \citenamefont {Pati}}]{Zhou:2018bon}%
  \BibitemOpen
  \bibfield  {author} {\bibinfo {author} {\bibfnamefont {S.}~\bibnamefont
  {Zhou}}, \bibinfo {author} {\bibfnamefont {P.}~\bibnamefont {Giuliani}},
  \bibinfo {author} {\bibfnamefont {J.}~\bibnamefont {Piekarewicz}}, \bibinfo
  {author} {\bibfnamefont {A.}~\bibnamefont {Bhattacharya}}, \ and\ \bibinfo
  {author} {\bibfnamefont {D.}~\bibnamefont {Pati}},\ }\href {\doibase
  10.1103/PhysRevC.99.055202} {\bibfield  {journal} {\bibinfo  {journal} {Phys.
  Rev.}\ }\textbf {\bibinfo {volume} {C99}},\ \bibinfo {pages} {055202}
  (\bibinfo {year} {2019})},\ \Eprint {http://arxiv.org/abs/1808.05977}
  {arXiv:1808.05977 [nucl-th]} \BibitemShut {NoStop}%
\bibitem [{\citenamefont {Mihovilovi\v{c}}\ \emph {et~al.}(2020)\citenamefont
  {Mihovilovi\v{c}}, \citenamefont {Higinbotham}, \citenamefont {Bevc},\ and\
  \citenamefont {\v{S}irca}}]{Mihovilovic:2020dmd}%
  \BibitemOpen
  \bibfield  {author} {\bibinfo {author} {\bibfnamefont {M.}~\bibnamefont
  {Mihovilovi\v{c}}}, \bibinfo {author} {\bibfnamefont {D.~W.}\ \bibnamefont
  {Higinbotham}}, \bibinfo {author} {\bibfnamefont {M.}~\bibnamefont {Bevc}}, \
  and\ \bibinfo {author} {\bibfnamefont {S.}~\bibnamefont {\v{S}irca}},\ }\href
  {\doibase 10.3389/fphy.2020.00036} {\bibfield  {journal} {\bibinfo  {journal}
  {Front. in Phys.}\ }\textbf {\bibinfo {volume} {8}},\ \bibinfo {pages} {36}
  (\bibinfo {year} {2020})},\ \Eprint {http://arxiv.org/abs/2003.03816}
  {arXiv:2003.03816 [nucl-ex]} \BibitemShut {NoStop}%
\bibitem [{\citenamefont {Bernauer}\ \emph {et~al.}(2010)\citenamefont
  {Bernauer} \emph {et~al.}}]{Bernauer:2010wm}%
  \BibitemOpen
  \bibfield  {author} {\bibinfo {author} {\bibfnamefont {J.~C.}\ \bibnamefont
  {Bernauer}} \emph {et~al.} (\bibinfo {collaboration} {A1}),\ }\href {\doibase
  10.1103/PhysRevLett.105.242001} {\bibfield  {journal} {\bibinfo  {journal}
  {Phys. Rev. Lett.}\ }\textbf {\bibinfo {volume} {105}},\ \bibinfo {pages}
  {242001} (\bibinfo {year} {2010})},\ \Eprint {http://arxiv.org/abs/1007.5076}
  {arXiv:1007.5076 [nucl-ex]} \BibitemShut {NoStop}%
\bibitem [{\citenamefont {Bernauer}\ \emph {et~al.}(2014)\citenamefont
  {Bernauer} \emph {et~al.}}]{Bernauer:2013tpr}%
  \BibitemOpen
  \bibfield  {author} {\bibinfo {author} {\bibfnamefont {J.~C.}\ \bibnamefont
  {Bernauer}} \emph {et~al.} (\bibinfo {collaboration} {A1}),\ }\href {\doibase
  10.1103/PhysRevC.90.015206} {\bibfield  {journal} {\bibinfo  {journal} {Phys.
  Rev.}\ }\textbf {\bibinfo {volume} {C90}},\ \bibinfo {pages} {015206}
  (\bibinfo {year} {2014})},\ \Eprint {http://arxiv.org/abs/1307.6227}
  {arXiv:1307.6227 [nucl-ex]} \BibitemShut {NoStop}%
\bibitem [{\citenamefont {{Merkle}}(2012)}]{Merkle:2012}%
  \BibitemOpen
  \bibfield  {author} {\bibinfo {author} {\bibfnamefont {M.}~\bibnamefont
  {{Merkle}}},\ }in\ \href@noop {} {\emph {\bibinfo {booktitle} {Analytic
  Number Theory, Approximation Theory, and Special Functions}}},\ \bibinfo
  {editor} {edited by\ \bibinfo {editor} {\bibfnamefont {G.~V.}\ \bibnamefont
  {Milovanovi\'{c}}}\ and\ \bibinfo {editor} {\bibfnamefont {M.~T.}\
  \bibnamefont {Rassias}}}\ (\bibinfo  {publisher} {Springer-Verlag New York},\
  \bibinfo {address} {New Your},\ \bibinfo {year} {2012})\ Chap.~\bibinfo
  {chapter} {12}, pp.\ \bibinfo {pages} {347--364},\ \Eprint
  {http://arxiv.org/abs/1211.0900} {arXiv:1211.0900} \BibitemShut {NoStop}%
\bibitem [{\citenamefont {Alarc\'{o}n}\ and\ \citenamefont
  {Weiss}(2018)}]{Alarcon:2018irp}%
  \BibitemOpen
  \bibfield  {author} {\bibinfo {author} {\bibfnamefont {J.~M.}\ \bibnamefont
  {Alarc\'{o}n}}\ and\ \bibinfo {author} {\bibfnamefont {C.}~\bibnamefont
  {Weiss}},\ }\href {\doibase 10.1016/j.physletb.2018.07.060} {\bibfield
  {journal} {\bibinfo  {journal} {Phys. Lett.}\ }\textbf {\bibinfo {volume}
  {B784}},\ \bibinfo {pages} {373} (\bibinfo {year} {2018})},\ \Eprint
  {http://arxiv.org/abs/1803.09748} {arXiv:1803.09748 [hep-ph]} \BibitemShut
  {NoStop}%
\bibitem [{\citenamefont {Press}\ \emph {et~al.}(2007)\citenamefont {Press},
  \citenamefont {Teukolsky}, \citenamefont {Vetterling},\ and\ \citenamefont
  {Flannery}}]{Recipes:2007}%
  \BibitemOpen
  \bibfield  {author} {\bibinfo {author} {\bibfnamefont {W.~H.}\ \bibnamefont
  {Press}}, \bibinfo {author} {\bibfnamefont {S.~A.}\ \bibnamefont
  {Teukolsky}}, \bibinfo {author} {\bibfnamefont {W.~T.}\ \bibnamefont
  {Vetterling}}, \ and\ \bibinfo {author} {\bibfnamefont {B.~P.}\ \bibnamefont
  {Flannery}},\ }\href@noop {} {\emph {\bibinfo {title} {Numerical Recipes 3rd
  Edition: The Art of Scientific Computing}}},\ \bibinfo {edition} {3rd}\ ed.\
  (\bibinfo  {publisher} {Cambridge University Press},\ \bibinfo {address} {New
  York, NY, USA},\ \bibinfo {year} {2007})\BibitemShut {NoStop}%
\bibitem [{\citenamefont {{\v{S}}irca}(2016)}]{Sirca:2016}%
  \BibitemOpen
  \bibfield  {author} {\bibinfo {author} {\bibfnamefont {S.}~\bibnamefont
  {{\v{S}}irca}},\ }\href {https://books.google.com/books?id=mqmPjwEACAAJ}
  {\emph {\bibinfo {title} {Probability for Physicists}}},\ Graduate Texts in
  Physics\ (\bibinfo  {publisher} {Springer International Publishing},\
  \bibinfo {year} {2016})\BibitemShut {NoStop}%
\bibitem [{\citenamefont {Akaike}(1974)}]{Akaike:1974}%
  \BibitemOpen
  \bibfield  {author} {\bibinfo {author} {\bibfnamefont {H.}~\bibnamefont
  {Akaike}},\ }\href {\doibase 10.1109/TAC.1974.1100705} {\bibfield  {journal}
  {\bibinfo  {journal} {IEEE Transactions on Automatic Control}\ }\textbf
  {\bibinfo {volume} {19}},\ \bibinfo {pages} {716} (\bibinfo {year}
  {1974})}\BibitemShut {NoStop}%
\bibitem [{\citenamefont {Schwarz}(1978)}]{Schwarz:1978}%
  \BibitemOpen
  \bibfield  {author} {\bibinfo {author} {\bibfnamefont {G.}~\bibnamefont
  {Schwarz}},\ }\href {\doibase 10.1214/aos/1176344136} {\bibfield  {journal}
  {\bibinfo  {journal} {Ann. Statist.}\ }\textbf {\bibinfo {volume} {6}},\
  \bibinfo {pages} {461} (\bibinfo {year} {1978})}\BibitemShut {NoStop}%
\bibitem [{\citenamefont {Higinbotham}\ \emph {et~al.}(2018)\citenamefont
  {Higinbotham}, \citenamefont {Giuliani}, \citenamefont {McClellan},
  \citenamefont {Sirca},\ and\ \citenamefont {Yan}}]{Higinbotham:2018jfh}%
  \BibitemOpen
  \bibfield  {author} {\bibinfo {author} {\bibfnamefont {D.~W.}\ \bibnamefont
  {Higinbotham}}, \bibinfo {author} {\bibfnamefont {P.}~\bibnamefont
  {Giuliani}}, \bibinfo {author} {\bibfnamefont {R.~E.}\ \bibnamefont
  {McClellan}}, \bibinfo {author} {\bibfnamefont {S.}~\bibnamefont {Sirca}}, \
  and\ \bibinfo {author} {\bibfnamefont {X.}~\bibnamefont {Yan}},\ }\href@noop
  {} {\  (\bibinfo {year} {2018})},\ \Eprint {http://arxiv.org/abs/1812.05706}
  {arXiv:1812.05706 [physics.data-an]} \BibitemShut {NoStop}%
\bibitem [{\citenamefont {{Andrae}}\ \emph {et~al.}(2010)\citenamefont
  {{Andrae}}, \citenamefont {{Schulze-Hartung}},\ and\ \citenamefont
  {{Melchior}}}]{Rene:2010}%
  \BibitemOpen
  \bibfield  {author} {\bibinfo {author} {\bibfnamefont {R.}~\bibnamefont
  {{Andrae}}}, \bibinfo {author} {\bibfnamefont {T.}~\bibnamefont
  {{Schulze-Hartung}}}, \ and\ \bibinfo {author} {\bibfnamefont
  {P.}~\bibnamefont {{Melchior}}},\ }\href@noop {} {\bibfield  {journal}
  {\bibinfo  {journal} {arXiv e-prints}\ ,\ \bibinfo {eid} {arXiv:1012.3754}}
  (\bibinfo {year} {2010})},\ \Eprint {http://arxiv.org/abs/1012.3754}
  {arXiv:1012.3754 [astro-ph.IM]} \BibitemShut {NoStop}%
\bibitem [{\citenamefont {Wit}\ \emph {et~al.}(2012)\citenamefont {Wit},
  \citenamefont {Heuvel},\ and\ \citenamefont {Romeijn}}]{Ernst:2012}%
  \BibitemOpen
  \bibfield  {author} {\bibinfo {author} {\bibfnamefont {E.}~\bibnamefont
  {Wit}}, \bibinfo {author} {\bibfnamefont {E.~v.~d.}\ \bibnamefont {Heuvel}},
  \ and\ \bibinfo {author} {\bibfnamefont {J.-W.}\ \bibnamefont {Romeijn}},\
  }\href {\doibase 10.1111/j.1467-9574.2012.00530.x} {\bibfield  {journal}
  {\bibinfo  {journal} {Statistica Nederlandica}\ }\textbf {\bibinfo {volume}
  {66}},\ \bibinfo {pages} {217} (\bibinfo {year} {2012})}\BibitemShut
  {NoStop}%
\bibitem [{\citenamefont {Anscombe}(1973)}]{Anscombe:1973}%
  \BibitemOpen
  \bibfield  {author} {\bibinfo {author} {\bibfnamefont {F.~J.}\ \bibnamefont
  {Anscombe}},\ }\href {http://www.jstor.org/stable/2682899} {\bibfield
  {journal} {\bibinfo  {journal} {The American Statistician}\ }\textbf
  {\bibinfo {volume} {27}},\ \bibinfo {pages} {17} (\bibinfo {year}
  {1973})}\BibitemShut {NoStop}%
\bibitem [{\citenamefont {Bezginov}\ \emph {et~al.}(2019)\citenamefont
  {Bezginov}, \citenamefont {Valdez}, \citenamefont {Horbatsch}, \citenamefont
  {Marsman}, \citenamefont {Vutha},\ and\ \citenamefont
  {Hessels}}]{Bezginov:2019}%
  \BibitemOpen
  \bibfield  {author} {\bibinfo {author} {\bibfnamefont {N.}~\bibnamefont
  {Bezginov}}, \bibinfo {author} {\bibfnamefont {T.}~\bibnamefont {Valdez}},
  \bibinfo {author} {\bibfnamefont {M.}~\bibnamefont {Horbatsch}}, \bibinfo
  {author} {\bibfnamefont {A.}~\bibnamefont {Marsman}}, \bibinfo {author}
  {\bibfnamefont {A.~C.}\ \bibnamefont {Vutha}}, \ and\ \bibinfo {author}
  {\bibfnamefont {E.~A.}\ \bibnamefont {Hessels}},\ }\href {\doibase
  10.1126/science.aau7807} {\bibfield  {journal} {\bibinfo  {journal}
  {Science}\ }\textbf {\bibinfo {volume} {365}},\ \bibinfo {pages} {1007}
  (\bibinfo {year} {2019})}\BibitemShut {NoStop}%
\bibitem [{\citenamefont {Koehler}(1993)}]{bias}%
  \BibitemOpen
  \bibfield  {author} {\bibinfo {author} {\bibfnamefont {J.~J.}\ \bibnamefont
  {Koehler}},\ }\href {\doibase https://doi.org/10.1006/obhd.1993.1044}
  {\bibfield  {journal} {\bibinfo  {journal} {Organizational Behavior and Human
  Decision Processes}\ }\textbf {\bibinfo {volume} {56}},\ \bibinfo {pages} {28
  } (\bibinfo {year} {1993})}\BibitemShut {NoStop}%
\bibitem [{\citenamefont {Ioannidis}(2005)}]{false}%
  \BibitemOpen
  \bibfield  {author} {\bibinfo {author} {\bibfnamefont {J.~P.~A.}\
  \bibnamefont {Ioannidis}},\ }\href {\doibase 10.1371/journal.pmed.0020124}
  {\bibfield  {journal} {\bibinfo  {journal} {PLOS Medicine}\ }\textbf
  {\bibinfo {volume} {2}} (\bibinfo {year} {2005}),\
  10.1371/journal.pmed.0020124}\BibitemShut {NoStop}%
\bibitem [{\citenamefont {Dyson}(2004)}]{Dyson:2004}%
  \BibitemOpen
  \bibfield  {author} {\bibinfo {author} {\bibfnamefont {F.}~\bibnamefont
  {Dyson}},\ }\href {http://dx.doi.org/10.1038/427297a} {\bibfield  {journal}
  {\bibinfo  {journal} {Nature}\ }\textbf {\bibinfo {volume} {427}},\ \bibinfo
  {pages} {297 EP } (\bibinfo {year} {2004})}\BibitemShut {NoStop}%
\bibitem [{\citenamefont {Yan}\ \emph {et~al.}(2018)\citenamefont {Yan},
  \citenamefont {Higinbotham}, \citenamefont {Dutta}, \citenamefont {Gao},
  \citenamefont {Gasparian}, \citenamefont {Khandaker}, \citenamefont
  {Liyanage}, \citenamefont {Pasyuk}, \citenamefont {Peng},\ and\ \citenamefont
  {Xiong}}]{Yan:2018bez}%
  \BibitemOpen
  \bibfield  {author} {\bibinfo {author} {\bibfnamefont {X.}~\bibnamefont
  {Yan}}, \bibinfo {author} {\bibfnamefont {D.~W.}\ \bibnamefont
  {Higinbotham}}, \bibinfo {author} {\bibfnamefont {D.}~\bibnamefont {Dutta}},
  \bibinfo {author} {\bibfnamefont {H.}~\bibnamefont {Gao}}, \bibinfo {author}
  {\bibfnamefont {A.}~\bibnamefont {Gasparian}}, \bibinfo {author}
  {\bibfnamefont {M.~A.}\ \bibnamefont {Khandaker}}, \bibinfo {author}
  {\bibfnamefont {N.}~\bibnamefont {Liyanage}}, \bibinfo {author}
  {\bibfnamefont {E.}~\bibnamefont {Pasyuk}}, \bibinfo {author} {\bibfnamefont
  {C.}~\bibnamefont {Peng}}, \ and\ \bibinfo {author} {\bibfnamefont
  {W.}~\bibnamefont {Xiong}},\ }\href {\doibase 10.1103/PhysRevC.98.025204}
  {\bibfield  {journal} {\bibinfo  {journal} {Phys. Rev.}\ }\textbf {\bibinfo
  {volume} {C98}},\ \bibinfo {pages} {025204} (\bibinfo {year} {2018})},\
  \Eprint {http://arxiv.org/abs/1803.01629} {arXiv:1803.01629 [nucl-ex]}
  \BibitemShut {NoStop}%
\bibitem [{\citenamefont {Xiong}\ \emph {et~al.}(2019)\citenamefont {Xiong}
  \emph {et~al.}}]{Xiong:2019umf}%
  \BibitemOpen
  \bibfield  {author} {\bibinfo {author} {\bibfnamefont {W.}~\bibnamefont
  {Xiong}} \emph {et~al.},\ }\href {\doibase 10.1038/s41586-019-1721-2}
  {\bibfield  {journal} {\bibinfo  {journal} {Nature}\ }\textbf {\bibinfo
  {volume} {575}},\ \bibinfo {pages} {147} (\bibinfo {year}
  {2019})}\BibitemShut {NoStop}%
\bibitem [{\citenamefont {Hand}\ \emph {et~al.}(1963)\citenamefont {Hand},
  \citenamefont {Miller},\ and\ \citenamefont {Wilson}}]{Hand:1963zz}%
  \BibitemOpen
  \bibfield  {author} {\bibinfo {author} {\bibfnamefont {L.~N.}\ \bibnamefont
  {Hand}}, \bibinfo {author} {\bibfnamefont {D.~G.}\ \bibnamefont {Miller}}, \
  and\ \bibinfo {author} {\bibfnamefont {R.}~\bibnamefont {Wilson}},\ }\href
  {\doibase 10.1103/RevModPhys.35.335} {\bibfield  {journal} {\bibinfo
  {journal} {Rev. Mod. Phys.}\ }\textbf {\bibinfo {volume} {35}},\ \bibinfo
  {pages} {335} (\bibinfo {year} {1963})}\BibitemShut {NoStop}%
\bibitem [{\citenamefont {Murphy}\ \emph {et~al.}(1974)\citenamefont {Murphy},
  \citenamefont {Shin},\ and\ \citenamefont {Skopik}}]{Murphy:1974zz}%
  \BibitemOpen
  \bibfield  {author} {\bibinfo {author} {\bibfnamefont {J.~J.}\ \bibnamefont
  {Murphy}}, \bibinfo {author} {\bibfnamefont {Y.~M.}\ \bibnamefont {Shin}}, \
  and\ \bibinfo {author} {\bibfnamefont {D.~M.}\ \bibnamefont {Skopik}},\
  }\href {\doibase 10.1103/PhysRevC.9.2125, 10.1103/PhysRevC.10.2111}
  {\bibfield  {journal} {\bibinfo  {journal} {Phys. Rev.}\ }\textbf {\bibinfo
  {volume} {C9}},\ \bibinfo {pages} {2125} (\bibinfo {year} {1974})},\ \bibinfo
  {note} {[Erratum: Phys. Rev. {\bf{C10}}, 2111 (1974)]}\BibitemShut {NoStop}%
\bibitem [{\citenamefont {Borkowski}\ \emph {et~al.}(1975)\citenamefont
  {Borkowski}, \citenamefont {Simon}, \citenamefont {Walther},\ and\
  \citenamefont {Wendling}}]{Borkowski:1975ume}%
  \BibitemOpen
  \bibfield  {author} {\bibinfo {author} {\bibfnamefont {F.}~\bibnamefont
  {Borkowski}}, \bibinfo {author} {\bibfnamefont {G.~G.}\ \bibnamefont
  {Simon}}, \bibinfo {author} {\bibfnamefont {V.~H.}\ \bibnamefont {Walther}},
  \ and\ \bibinfo {author} {\bibfnamefont {R.~D.}\ \bibnamefont {Wendling}},\
  }\href {\doibase 10.1007/BF01409496} {\bibfield  {journal} {\bibinfo
  {journal} {Z. Phys.}\ }\textbf {\bibinfo {volume} {A275}},\ \bibinfo {pages}
  {29} (\bibinfo {year} {1975})}\BibitemShut {NoStop}%
\bibitem [{\citenamefont {Simon}\ \emph {et~al.}(1980)\citenamefont {Simon},
  \citenamefont {Schmitt}, \citenamefont {Borkowski},\ and\ \citenamefont
  {Walther}}]{Simon:1980hu}%
  \BibitemOpen
  \bibfield  {author} {\bibinfo {author} {\bibfnamefont {G.~G.}\ \bibnamefont
  {Simon}}, \bibinfo {author} {\bibfnamefont {C.}~\bibnamefont {Schmitt}},
  \bibinfo {author} {\bibfnamefont {F.}~\bibnamefont {Borkowski}}, \ and\
  \bibinfo {author} {\bibfnamefont {V.~H.}\ \bibnamefont {Walther}},\ }\href
  {\doibase 10.1016/0375-9474(80)90104-9} {\bibfield  {journal} {\bibinfo
  {journal} {Nucl. Phys.}\ }\textbf {\bibinfo {volume} {A333}},\ \bibinfo
  {pages} {381} (\bibinfo {year} {1980})}\BibitemShut {NoStop}%
\bibitem [{\citenamefont {Hagelstein}\ and\ \citenamefont
  {Pascalutsa}(2019)}]{Hagelstein:2018zrz}%
  \BibitemOpen
  \bibfield  {author} {\bibinfo {author} {\bibfnamefont {F.}~\bibnamefont
  {Hagelstein}}\ and\ \bibinfo {author} {\bibfnamefont {V.}~\bibnamefont
  {Pascalutsa}},\ }\href {\doibase 10.1016/j.physletb.2019.134825} {\bibfield
  {journal} {\bibinfo  {journal} {Phys. Lett.}\ }\textbf {\bibinfo {volume}
  {B797}},\ \bibinfo {pages} {134825} (\bibinfo {year} {2019})},\ \Eprint
  {http://arxiv.org/abs/1812.02028} {arXiv:1812.02028 [nucl-th]} \BibitemShut
  {NoStop}%
\bibitem [{\citenamefont {Hayward}\ and\ \citenamefont
  {Griffioen}(2018)}]{Hayward:2018qij}%
  \BibitemOpen
  \bibfield  {author} {\bibinfo {author} {\bibfnamefont {T.~B.}\ \bibnamefont
  {Hayward}}\ and\ \bibinfo {author} {\bibfnamefont {K.~A.}\ \bibnamefont
  {Griffioen}},\ }\href@noop {} {\  (\bibinfo {year} {2018})},\ \Eprint
  {http://arxiv.org/abs/1804.09150} {arXiv:1804.09150 [nucl-ex]} \BibitemShut
  {NoStop}%
\bibitem [{\citenamefont {Bernard}\ \emph {et~al.}(1998)\citenamefont
  {Bernard}, \citenamefont {Fearing}, \citenamefont {Hemmert},\ and\
  \citenamefont {Mei{\ss}ner}}]{Bernard:1998gv}%
  \BibitemOpen
  \bibfield  {author} {\bibinfo {author} {\bibfnamefont {V.}~\bibnamefont
  {Bernard}}, \bibinfo {author} {\bibfnamefont {H.~W.}\ \bibnamefont
  {Fearing}}, \bibinfo {author} {\bibfnamefont {T.~R.}\ \bibnamefont
  {Hemmert}}, \ and\ \bibinfo {author} {\bibfnamefont {{\relax Ulf}.-G.}\
  \bibnamefont {Mei{\ss}ner}},\ }\href {\doibase 10.1016/S0375-9474(98)00175-4,
  10.1016/S0375-9474(98)00566-1} {\bibfield  {journal} {\bibinfo  {journal}
  {Nucl. Phys.}\ }\textbf {\bibinfo {volume} {A635}},\ \bibinfo {pages} {121}
  (\bibinfo {year} {1998})},\ \bibinfo {note} {[Erratum: Nucl.
  Phys.A642,563(1998)]},\ \Eprint {http://arxiv.org/abs/hep-ph/9801297}
  {arXiv:hep-ph/9801297 [hep-ph]} \BibitemShut {NoStop}%
\bibitem [{\citenamefont {Hohler}\ \emph {et~al.}(1976)\citenamefont {Hohler},
  \citenamefont {Pietarinen}, \citenamefont {Sabba~Stefanescu}, \citenamefont
  {Borkowski}, \citenamefont {Simon}, \citenamefont {Walther},\ and\
  \citenamefont {Wendling}}]{Hohler:1976ax}%
  \BibitemOpen
  \bibfield  {author} {\bibinfo {author} {\bibfnamefont {G.}~\bibnamefont
  {Hohler}}, \bibinfo {author} {\bibfnamefont {E.}~\bibnamefont {Pietarinen}},
  \bibinfo {author} {\bibfnamefont {I.}~\bibnamefont {Sabba~Stefanescu}},
  \bibinfo {author} {\bibfnamefont {F.}~\bibnamefont {Borkowski}}, \bibinfo
  {author} {\bibfnamefont {G.~G.}\ \bibnamefont {Simon}}, \bibinfo {author}
  {\bibfnamefont {V.~H.}\ \bibnamefont {Walther}}, \ and\ \bibinfo {author}
  {\bibfnamefont {R.~D.}\ \bibnamefont {Wendling}},\ }\href {\doibase
  10.1016/0550-3213(76)90449-1} {\bibfield  {journal} {\bibinfo  {journal}
  {Nucl. Phys.}\ }\textbf {\bibinfo {volume} {B114}},\ \bibinfo {pages} {505}
  (\bibinfo {year} {1976})}\BibitemShut {NoStop}%
\bibitem [{\citenamefont {Lorenz}\ \emph {et~al.}(2015)\citenamefont {Lorenz},
  \citenamefont {Mei{\ss}ner}, \citenamefont {Hammer},\ and\ \citenamefont
  {Dong}}]{Lorenz:2014yda}%
  \BibitemOpen
  \bibfield  {author} {\bibinfo {author} {\bibfnamefont {I.~T.}\ \bibnamefont
  {Lorenz}}, \bibinfo {author} {\bibfnamefont {{\relax Ulf}.-G.}\ \bibnamefont
  {Mei{\ss}ner}}, \bibinfo {author} {\bibfnamefont {H.~W.}\ \bibnamefont
  {Hammer}}, \ and\ \bibinfo {author} {\bibfnamefont {Y.~B.}\ \bibnamefont
  {Dong}},\ }\href {\doibase 10.1103/PhysRevD.91.014023} {\bibfield  {journal}
  {\bibinfo  {journal} {Phys. Rev.}\ }\textbf {\bibinfo {volume} {D91}},\
  \bibinfo {pages} {014023} (\bibinfo {year} {2015})},\ \Eprint
  {http://arxiv.org/abs/1411.1704} {arXiv:1411.1704 [hep-ph]} \BibitemShut
  {NoStop}%
\bibitem [{\citenamefont {Kelly}(2004)}]{Kelly:2004hm}%
  \BibitemOpen
  \bibfield  {author} {\bibinfo {author} {\bibfnamefont {J.~J.}\ \bibnamefont
  {Kelly}},\ }\href {\doibase 10.1103/PhysRevC.70.068202} {\bibfield  {journal}
  {\bibinfo  {journal} {Phys. Rev.}\ }\textbf {\bibinfo {volume} {C70}},\
  \bibinfo {pages} {068202} (\bibinfo {year} {2004})}\BibitemShut {NoStop}%
\bibitem [{\citenamefont {Sick}(2003)}]{Sick:2003gm}%
  \BibitemOpen
  \bibfield  {author} {\bibinfo {author} {\bibfnamefont {I.}~\bibnamefont
  {Sick}},\ }\href {\doibase 10.1016/j.physletb.2003.09.092} {\bibfield
  {journal} {\bibinfo  {journal} {Phys. Lett.}\ }\textbf {\bibinfo {volume}
  {B576}},\ \bibinfo {pages} {62} (\bibinfo {year} {2003})},\ \Eprint
  {http://arxiv.org/abs/nucl-ex/0310008} {arXiv:nucl-ex/0310008 [nucl-ex]}
  \BibitemShut {NoStop}%
\bibitem [{\citenamefont {Lorenz}\ \emph {et~al.}(2012)\citenamefont {Lorenz},
  \citenamefont {Hammer},\ and\ \citenamefont {Mei{\ss}ner}}]{Lorenz:2012tm}%
  \BibitemOpen
  \bibfield  {author} {\bibinfo {author} {\bibfnamefont {I.~T.}\ \bibnamefont
  {Lorenz}}, \bibinfo {author} {\bibfnamefont {H.~W.}\ \bibnamefont {Hammer}},
  \ and\ \bibinfo {author} {\bibfnamefont {{\relax Ulf}.-G.}\ \bibnamefont
  {Mei{\ss}ner}},\ }\href {\doibase 10.1140/epja/i2012-12151-1} {\bibfield
  {journal} {\bibinfo  {journal} {Eur. Phys. J.}\ }\textbf {\bibinfo {volume}
  {A48}},\ \bibinfo {pages} {151} (\bibinfo {year} {2012})},\ \Eprint
  {http://arxiv.org/abs/1205.6628} {arXiv:1205.6628 [hep-ph]} \BibitemShut
  {NoStop}%
\bibitem [{\citenamefont {Hill}\ and\ \citenamefont {Paz}(2010)}]{Hill:2010yb}%
  \BibitemOpen
  \bibfield  {author} {\bibinfo {author} {\bibfnamefont {R.~J.}\ \bibnamefont
  {Hill}}\ and\ \bibinfo {author} {\bibfnamefont {G.}~\bibnamefont {Paz}},\
  }\href {\doibase 10.1103/PhysRevD.82.113005} {\bibfield  {journal} {\bibinfo
  {journal} {Phys. Rev.}\ }\textbf {\bibinfo {volume} {D82}},\ \bibinfo {pages}
  {113005} (\bibinfo {year} {2010})},\ \Eprint {http://arxiv.org/abs/1008.4619}
  {arXiv:1008.4619 [hep-ph]} \BibitemShut {NoStop}%
\bibitem [{\citenamefont {Higinbotham}(2019)}]{Higinbotham-Code-2019}%
  \BibitemOpen
  \bibfield  {author} {\bibinfo {author} {\bibfnamefont {D.~W.}\ \bibnamefont
  {Higinbotham}},\ }\href {\doibase 10.5281/zenodo.2566639} {\  (\bibinfo
  {year} {2019}),\ 10.5281/zenodo.2566639}\BibitemShut {NoStop}%
\bibitem [{\citenamefont {Draper}(1988)}]{Draper:robust}%
  \BibitemOpen
  \bibfield  {author} {\bibinfo {author} {\bibfnamefont {D.}~\bibnamefont
  {Draper}},\ }\href {\doibase doi:10.1214/ss/1177012915} {\bibfield  {journal}
  {\bibinfo  {journal} {Statistical Science}\ }\textbf {\bibinfo {volume}
  {3}},\ \bibinfo {pages} {239} (\bibinfo {year} {1988})}\BibitemShut {NoStop}%
\bibitem [{\citenamefont {Triggs}\ \emph {et~al.}(2000)\citenamefont {Triggs},
  \citenamefont {McLauchlan}, \citenamefont {Hartley},\ and\ \citenamefont
  {Fitzgibbon}}]{triggs:2000}%
  \BibitemOpen
  \bibfield  {author} {\bibinfo {author} {\bibfnamefont {B.}~\bibnamefont
  {Triggs}}, \bibinfo {author} {\bibfnamefont {P.~F.}\ \bibnamefont
  {McLauchlan}}, \bibinfo {author} {\bibfnamefont {R.~I.}\ \bibnamefont
  {Hartley}}, \ and\ \bibinfo {author} {\bibfnamefont {A.~W.}\ \bibnamefont
  {Fitzgibbon}},\ }in\ \href {\doibase 10.1007/3-540-44480-7_21} {\emph
  {\bibinfo {booktitle} {Vision Algorithms: Theory and Practice}}},\ \bibinfo
  {editor} {edited by\ \bibinfo {editor} {\bibfnamefont {B.}~\bibnamefont
  {Triggs}}, \bibinfo {editor} {\bibfnamefont {A.}~\bibnamefont {Zisserman}}, \
  and\ \bibinfo {editor} {\bibfnamefont {R.}~\bibnamefont {Szeliski}}}\
  (\bibinfo  {publisher} {Springer Berlin Heidelberg},\ \bibinfo {address}
  {Berlin, Heidelberg},\ \bibinfo {year} {2000})\ pp.\ \bibinfo {pages}
  {298--372}\BibitemShut {NoStop}%
\bibitem [{\citenamefont {Jones}\ \emph {et~al.}(01  )\citenamefont {Jones},
  \citenamefont {Oliphant}, \citenamefont {Peterson} \emph
  {et~al.}}]{Oliphant:2001}%
  \BibitemOpen
  \bibfield  {author} {\bibinfo {author} {\bibfnamefont {E.}~\bibnamefont
  {Jones}}, \bibinfo {author} {\bibfnamefont {T.}~\bibnamefont {Oliphant}},
  \bibinfo {author} {\bibfnamefont {P.}~\bibnamefont {Peterson}},  \emph
  {et~al.},\ }\href {http://www.scipy.org/} {\enquote {\bibinfo {title}
  {{SciPy}: Open source scientific tools for {Python}},}\ } (\bibinfo {year}
  {2001--}),\ \bibinfo {note} {[Online; accessed 9/26/2019]}\BibitemShut
  {NoStop}%
\bibitem [{\citenamefont {Oliphant}(2007)}]{Oliphant:2007}%
  \BibitemOpen
  \bibfield  {author} {\bibinfo {author} {\bibfnamefont {T.~E.}\ \bibnamefont
  {Oliphant}},\ }\href {\doibase 10.1109/MCSE.2007.58} {\bibfield  {journal}
  {\bibinfo  {journal} {Computing in Science $\&$ Engineering}\ }\textbf
  {\bibinfo {volume} {9}},\ \bibinfo {pages} {10} (\bibinfo {year}
  {2007})}\BibitemShut {NoStop}%
\bibitem [{\citenamefont {Mihovilovič}\ \emph {et~al.}(2019)\citenamefont
  {Mihovilovič} \emph {et~al.}}]{Mihovilovic:2019jiz}%
  \BibitemOpen
  \bibfield  {author} {\bibinfo {author} {\bibfnamefont {M.}~\bibnamefont
  {Mihovilovič}} \emph {et~al.},\ }\href@noop {} {\  (\bibinfo {year}
  {2019})},\ \Eprint {http://arxiv.org/abs/1905.11182} {arXiv:1905.11182
  [nucl-ex]} \BibitemShut {NoStop}%
\bibitem [{\citenamefont {Mihovilovi\v{c}}\ \emph {et~al.}(2017)\citenamefont
  {Mihovilovi\v{c}} \emph {et~al.}}]{Mihovilovic:2016rkr}%
  \BibitemOpen
  \bibfield  {author} {\bibinfo {author} {\bibfnamefont {M.}~\bibnamefont
  {Mihovilovi\v{c}}} \emph {et~al.},\ }\href {\doibase
  10.1016/j.physletb.2017.05.031} {\bibfield  {journal} {\bibinfo  {journal}
  {Phys. Lett.}\ }\textbf {\bibinfo {volume} {B771}},\ \bibinfo {pages} {194}
  (\bibinfo {year} {2017})},\ \Eprint {http://arxiv.org/abs/1612.06707}
  {arXiv:1612.06707 [nucl-ex]} \BibitemShut {NoStop}%
\bibitem [{\citenamefont {\v{S}irca}\ and\ \citenamefont
  {Horvat}(2012)}]{Sirca:2012}%
  \BibitemOpen
  \bibfield  {author} {\bibinfo {author} {\bibfnamefont {S.}~\bibnamefont
  {\v{S}irca}}\ and\ \bibinfo {author} {\bibfnamefont {M.}~\bibnamefont
  {Horvat}},\ }\href@noop {} {\emph {\bibinfo {title} {Computational Methods
  for Physicists}}}\ (\bibinfo  {publisher} {Springer},\ \bibinfo {year}
  {2012})\BibitemShut {NoStop}%
\end{thebibliography}

%
\end{document}